\documentclass[10pt,a4paper,twocolumn]{article}
\usepackage[top=20mm,bottom=20mm,left=15mm,right=15mm,columnsep=6mm]{geometry}
\usepackage{amsmath,amssymb,bm,graphicx,booktabs,array}
\usepackage{fontspec,xeCJK}
\usepackage{lmodern}
\usepackage{microtype}
\usepackage[numbers,sort&compress]{natbib}
\usepackage{cuted}
\newenvironment{widetext}{\begin{strip}}{\end{strip}}
\usepackage[colorlinks=true,linkcolor=blue,citecolor=blue,urlcolor=blue]{hyperref}
\newcommand{\Tr}{\operatorname{Tr}}
\newcommand{\dd}{\mathrm{d}}
\newcommand{\Nc}{N_c}
\newcommand{\CF}{C_F}
\newcommand{\qb}{\bar z}
\newcommand{\eps}{\epsilon}
\newcommand{\brho}{\bar\rho}
\newcommand{\bR}{\bar R}
\newcommand{\cN}{\mathcal N}
\newcommand{\fnorm}[1]{\left\|#1\right\|_{\mathrm F}}
\newcommand{\avg}[1]{\left\langle #1\right\rangle_\lambda}
\newcommand{\uv}[1]{\bm{#1}}
\begin{document}
\twocolumn[{\begin{@twocolumnfalse}
\begin{center}
{\Large\bfseries Power corrections to heavy-quark spin correlations in small-$x$ deep-inelastic scattering\textsuperscript{*}\par}
\vspace{8pt}
{\large Meisen Gao (高梅森)\textsuperscript{1,2,\textdagger}}\\[5pt]
{\small\textsuperscript{1}School of Physics, East China University of Science and Technology, Shanghai 200237, China\\
\textsuperscript{2}Shanghai Key Laboratory of Particle Physics and Cosmology, Shanghai 200240, China}
\end{center}
\begin{abstract}
Heavy-quark spin correlations in small-$x$ deep-inelastic scattering receive corrections beyond the back-to-back limit. For a rotationally symmetric target at leading power, the target dependence cancels from the normalized spin density matrix after full azimuthal averaging, for fixed kinematics and spin axes related by transverse light-front boosts. We calculate the first nonvanishing correction at tree level for finite quark mass and finite $N_c$ in a homogeneous Gaussian target. The expansion through fourth order in the dipole sizes includes logarithms, contact terms and all interference contributions at that order. Comparison with the improved transverse-momentum-dependent approximation separates kinematic corrections from terms with additional insertions of the target current. Numerical integration of the unexpanded Gaussian kernel tests the approximation at 28 parameter points. Spin correlations differ between the two targets in all 12 kinematic comparisons; in one comparison both spin states remain separable. At fixed photon virtuality, the leading large-$P$ behavior of the fourth-order correction to the normalized transverse spin state is generically proportional to $m\ln P/P^3$. We also compare two fixed Gaussian targets using unpolarized momentum distributions and joint spin measurements. For a purely transverse photon, three narrow momentum bins and an ideal measurement of both spins along the common quark-momentum axis defined by the light-front boost, spin information adds $3.5$--$3.7\%$ to the expected log likelihood ratio after the overall normalization is fitted. This increase depends on the spin analyzing power and bin selection. The results quantify a model-dependent effect under idealized measurement conditions.
\end{abstract}
\vspace{12pt}
\end{@twocolumnfalse}}]
\begingroup
\renewcommand{\thefootnote}{\fnsymbol{footnote}}
\footnotetext[1]{Supported by the National Natural Science Foundation of China (Grant No. 12605224).}
\footnotetext[2]{Corresponding author: \href{mailto:msgao@ecust.edu.cn}{msgao@ecust.edu.cn}; ORCID: \href{https://orcid.org/0000-0002-4028-9895}{0000-0002-4028-9895}.}
\endgroup
\section{Introduction}
Heavy-quark pairs in deep-inelastic scattering (DIS) probe the transverse structure of small-$x$ gluons. The Color Glass Condensate (CGC) describes their coherent multiple scattering through Wilson-line correlators \cite{Gelis:2010nm}, within the saturation program for present and future electron--ion collisions \cite{Accardi:2012qut,Morreale:2021pnn,Aschenauer:2017jsk}. In the back-to-back, or correlation, limit, the inclusive pair cross section reduces to unpolarized and linearly polarized Weizs\"acker--Williams (WW) gluon distributions \cite{Dominguez2011}. Their tensor structure is part of the general classification of gluon transverse-momentum-dependent distributions (TMDs) \cite{Mulders:2000sh}. Heavy-pair azimuthal observables have long been proposed as probes of linear gluon polarization \cite{Boer:2010zf,Pisano:2013cya,Boer:2016fqd,Efremov:2017iwh}.

The Electron--Ion Collider in China (EicC) program emphasizes sea-quark and moderate-$x$ nucleon structure and nuclear effects \cite{Anderle:2021wcy,Xiao:2026tbs}. For the US EIC, Monte Carlo studies have compared open-charm, charged-dihadron and dijet channels for measuring the gluon Sivers function with a transversely polarized target \cite{Zheng:2018ssm}. Simulations of reconstructed and tagged heavy-flavor hadron pairs have also estimated the statistical precision of gluon-TMD azimuthal asymmetries \cite{Dong:2022xbd}. These studies concern momentum correlations; reconstructing the final-state spin correlations considered here requires additional spin-analyzing information.

Retaining the final quark spins gives a different set of observables. Collider spin tomography and entanglement measurements \cite{Afik:2020onf,ATLAS:2023fsd,CMS:2024pts}, reviewed in Ref.~\cite{Barr:2024djo}, motivate calculations of the full density matrix rather than only its trace. For top pairs, decay-angle observables project the independent spin coefficients \cite{Bernreuther:2015yna}, and predictions retain their invariant-mass and scattering-angle dependence \cite{Bernreuther:2024ltu}. For hadronized charm or bottom pairs, the spin response requires a separate description. DIS proposals now include generalized-parton-distribution and lepton--quark quantum correlations \cite{Hatta:2025obw,Cheng:2025zaw,Bloss:2026yrf}. Dihadron helicity correlations in photon--nucleus collisions have also been studied with collinear hard scattering and polarized fragmentation \cite{Chen:2024qvx}. The inclusive small-$x$ heavy-quark spin matrix of Ref.~\cite{Fucilla2026} addresses the correlation limit. At fixed remaining kinematics, with a prescribed identification of spin axes between events, full relative-azimuth averaging leaves a common WW factor that cancels from the normalized state.

We determine how the first surviving correction changes that state. A state change need not change entanglement, and a target-dependent spin observable need not add much discrimination once the unpolarized kinematics are used. These distinctions require both a calculation with explicit spin indices and a specified statistical comparison. They also require separating kinematic corrections multiplying the WW operator from contributions with additional insertions of the target current.

The ingredients of the amplitude expansion are established. The dipole--quadrupole production kernel is given in Ref.~\cite{Dominguez2011}, and the complete quartic kernel, including the $13$, $31$ and $22$ sectors and self-dipole logarithms, was obtained in the Gaussian large-$N_c$ treatment of Ref.~\cite{Dumitru2016}. Improved-TMD (ITMD) factorization \cite{Kotko:2015ura}, all-kinematic-twist resummation \cite{Altinoluk2019,Altinoluk:2019wyu}, and the transverse-link construction \cite{Boussarie:2020vzf} organize kinematic and genuine higher twists before the final spin sum. The same organization accommodates finite masses. Massive ITMD hard factors are available \cite{Altinoluk2021}, while Ref.~\cite{Boussarie2021} retains final helicities in its massless amplitudes before forming numerical spin-summed cross sections. Massive amplitudes with explicit quark spins and finite-$N_c$ Gaussian correlators also enter the CGC--NRQCD calculation of Ref.~\cite{Cheung:2024qvw}, where the quark pair is projected onto quarkonium channels. Recent operator approaches address the eikonal--TMD relation with open Dirac structure \cite{KarMethods2026,Kar2026} and arbitrary-$x$ twist-three terms \cite{Mukherjee:2026cte}. Finite-energy corrections \cite{Altinoluk:2024zom,Agostini:2025vvx} expand a different parameter from the transverse-size correction considered here. Ref.~\cite{Caucal:2026yzl} reviews the current small-$x$ TMD framework.

We evaluate the quartic correction to the normalized finite-mass pair-spin state in a homogeneous Gaussian target at finite $N_c$. Equation~\eqref{eq:k4} extends the Gaussian coordinate kernel of Ref.~\cite{Dumitru2016}; Appendix~\ref{app:source} specifies its symmetric-center large-$N_c$ reduction. Contracting the kernel with the massive spin amplitudes gives Eq.~\eqref{eq:r4}, including the self-dipole logarithms, contact terms and all same-order interference terms. We expand the normalization at the same order. Integrations of the unexpanded kernel test the approximation. The numerical comparison finds target-dependent spin correlations even for a pair of separable states. Longitudinal maximal entanglement is already known in photon--gluon fusion \cite{Qi2026}, diffraction \cite{FucillaDiffractive2026}, and the inclusive correlation limit \cite{Fucilla2026}; we formulate its cancellation for an arbitrary spin-independent inclusive eikonal target trace. General polarization and concurrence constraints \cite{Liu:2026dzv} provide a separate quantum-information context. For transverse photons, normalization cancels the leading common spin structure of the quartic term. The surviving mass-interference coefficient is given in Eq.~\eqref{eq:largep} at fixed $m,Q^2,z,q$ and target parameters.

Finally, we use the full-kernel matrices to evaluate the spin information added to a common unpolarized shape likelihood after profiling one overall normalization. For the primary three-bin comparison, an ideal measurement along the common quark-momentum axis defined by the light-front boost in the transverse-photon channel adds about $3.5$--$3.7\%$ to the expected log-likelihood information. The comparison uses two fixed Gaussian hypotheses and a specified measurement. Extending it to a continuous target fit or an EIC forecast requires further physical inputs. Gaussian closure relates all multipoint correlators to one covariance, so these correlators are not independently varied here. Fragmentation-based tomography \cite{FucillaTomography2026} and decoherence from unresolved next-to-soft radiation \cite{Agrawal2026} address further ingredients needed to connect the Born spin matrix to a measurement.

\section{Spin density matrix and production kernel}\label{sec:spin}
\subsection{Kinematics and the spin density matrix}
We consider $\gamma^* A\to Q\bar Q X$ at fixed small $x$, tree-level photon splitting and eikonal scattering, per unit transverse target area. With $p^\pm=(p^0\pm p^3)/\sqrt2$ and metric $(+---)$, define
\begin{align}
 z&=p_Q^+/p_\gamma^+,\qquad \qb=1-z,\nonumber\\
 \uv q&=\uv p_Q+\uv p_{\bar Q},\qquad
 \uv P=\qb\uv p_Q-z\uv p_{\bar Q},\nonumber\\
 \uv p_Q&=\uv P+z\uv q,\qquad
 \uv p_{\bar Q}=-\uv P+\qb\uv q,\nonumber\\
 \eps^2&=m^2+z\qb Q^2,\qquad D_P=P^2+\eps^2 .
 \label{eq:kin}
\end{align}
The conjugate coordinates are $\uv u=\uv x_1-\uv x_2$ and $\uv v=z\uv x_1+\qb\uv x_2$. Their Jacobian is unity and the phase of the squared amplitude is $P\cdot(u-u')+q\cdot(v-v')$. For $z\ne1/2$, $P$ must not be replaced by $(p_Q-p_{\bar Q})/2$.

We order the two-particle spin states as $(++,+-,-+,--)$, with
\begin{equation}
 R_{s\bar s;t\bar t}=\avg{\left\langle
 \mathcal A_{s\bar s}\mathcal A^*_{t\bar t}\right\rangle_A},
 \qquad \rho=R/\Tr R.
 \label{eq:density}
\end{equation}
Color is summed and averaged through the kernel below. The transverse photon average contains two orthogonal Cartesian polarizations with weight $1/2$; the longitudinal photon has weight one. The inclusive target average acts on the product of the amplitude and its conjugate.

To define the state after angular averaging, spin spaces at different momenta must be identified. We take $\uv P=(P,0)$ and define axes in the $q=0$ pair rest frame:
\begin{align}
 M^2&=\frac{P^2+m^2}{z\qb},\nonumber\\
 p_Q&=(M/2,P,0,(2z-1)M/2),\nonumber\\
 \hat k&=\boldsymbol p_Q/|\boldsymbol p_Q|,\quad
 \hat n=(0,1,0),\quad \hat r=\hat k\times\hat n .
 \label{eq:axes}
\end{align}
The Pauli axes $(x,y,z)$ label $(n,r,k)$. Both particles use the common $k$ axis; the antiquark phase convention is specified in Appendix~\ref{app:spin}. These rest-frame spins are transported to finite $q$ by the exact transverse light-front boost. Writing $b_i=q_i/p_\gamma^+$, the momentum components before and after the boost satisfy
\begin{align}
 p^{\prime +}&=p^+,\qquad p_i^\prime=p_i+b_i p^+,\nonumber\\
 p^{\prime -}&=p^-+b_i p_i+\tfrac12 b^2p^+,\nonumber\\
 S(b)&=1-\tfrac12 b_i\gamma^i\gamma^+ .
 \label{eq:boost}
\end{align}
Here $i=1,2$, repeated transverse indices are summed, $b^2=b_1^2+b_2^2$, and $S(b)$ acts on the Dirac spinors. The boosted external spinors obey the Dirac equations and completeness relations. Their inverse transformation returns the pair to its reference rest frame. At finite $q$, the photon direction in the reference rest frame differs from the reference $z$ axis. The boosted light-front basis therefore differs from a helicity basis defined by the photon direction in each event.

Let $\phi=\arg q-\arg P$. With all other variables fixed, the observable studied here is
\begin{equation}
 \bR=\frac1{2\pi}\int_0^{2\pi}\dd\phi\,R(\phi),
 \qquad \brho=\frac{\bR}{\Tr\bR}.
 \label{eq:average}
\end{equation}
In the basis defined by Eq.~\eqref{eq:boost}, no additional spin rotation is required under the angular integral. A different event-axis convention requires rotating each unnormalized matrix before integration. Uniformly averaging $\rho(\phi)$ or its negativity would define a different observable.

We report the complete correlations $C_{ij}=\Tr(\rho\,\sigma_i\otimes\sigma_j)$, the smallest eigenvalue of the antiquark partial transpose, and the negativity
\begin{equation}
 \cN(\rho)=\frac{\|\rho^{T_{\bar Q}}\|_1-1}{2}.
 \label{eq:neg}
\end{equation}
Here $T_{\bar Q}$ denotes partial transposition in the antiquark spin space, and $\|X\|_1=\Tr\sqrt{X^\dagger X}$ is the trace norm. The negativity is evaluated after normalization \cite{Peres1996,Vidal2002}. For two qubits, vanishing negativity is equivalent to separability \cite{Horodecki1996}. We also evaluate the general two-qubit concurrence \cite{Wootters1998} as a check, without imposing a special form on the density matrix.

The relation to spin observables follows from the Pauli decomposition
\begin{align}
 \rho=\frac14\Big[&I\otimes I+a_i\sigma_i\otimes I
 +b_j I\otimes\sigma_j\nonumber\\
 &+C_{ij}\sigma_i\otimes\sigma_j\Big].
\end{align}
For two normalized states at the same kinematics, define $\Delta\rho=\rho_B-\rho_A$ and the Frobenius norm $\|X\|_{\mathrm F}^2=\Tr(X^\dagger X)$. Orthogonality of the Pauli basis gives
\begin{equation}
 \fnorm{\Delta\rho}^2=\frac14\left(
 |\Delta a|^2+|\Delta b|^2+\sum_{ij}|\Delta C_{ij}|^2\right).
 \label{eq:paulidistance}
\end{equation}
Thus a state distance can be resolved into local polarizations and
two-particle spin correlations, without interpreting it as an entanglement
measure. The vectors $a,b$ in this decomposition are Pauli expectation
values and are distinct from the transverse coordinates below.

\subsection{Photon wave functions with explicit spin indices}
The photon wave functions with finite quark mass and explicit spin indices in the convention of Ref.~\cite{Fucilla2026} can be written
\begin{align}
 \Phi^L_a(k)&=\frac{z\qb Q}{p_\gamma^+}
 \frac{(\bar u\gamma^+v)_a}{k^2+\eps^2},\label{eq:wavefunctionL}\\
 \Phi^{T\lambda}_a(k)&=\frac{1}{2p_\gamma^+}
 \frac{\big[\bar u\{(1-2z)k_\lambda+
 i\epsilon^{\lambda j}k_j\gamma_5-m\gamma^\lambda\}\gamma^+v\big]_a}
 {k^2+\eps^2}.
 \label{eq:wavefunction}
\end{align}
Here $a=(s,\bar s)$ and $\epsilon^{12}=+1$. In all Fourier derivatives, the external on-shell momenta and spinors are held fixed; only the internal variable $k$ is differentiated. Setting $k=P$ is the last step. Differentiating the external spin basis or $M(P)$ would change this operation.

With $\Phi(u)=\int\dd^2k\,e^{iku}\Phi(k)/(2\pi)^2$, the amplitude contains
\begin{align}
 \mathcal D(u,v)&=U(v+\qb u)U^\dagger(v-zu)-1,\nonumber\\
 \mathcal A_a&\propto\int\dd^2u\,\dd^2v\,e^{-iPu-iqv}\Phi_a(u)\mathcal D(u,v).
\end{align}
The resulting inclusive kernel is
\begin{align}
 K(u,w,b)&=1+Q(x_1,x_2,x'_2,x'_1)\nonumber\\
 &\quad-S(x_1,x_2)-S(x'_2,x'_1),\nonumber\\
 Q(1,2,3,4)&=\frac1{\Nc}\left\langle\Tr[U_1U_2^\dagger U_3U_4^\dagger]\right\rangle_A.
 \label{eq:kernel}
\end{align}
Here $b=v-v'$ and $w=u'$.
Thus $K(0,w,b)=K(u,0,b)=0$. The two wave functions remain in a coherent double integral. Multiplying the unpolarized rate by the leading-power spin state would omit corrections to the spin dependence of this expression.

\subsection{Finite-\texorpdfstring{$N_c$}{Nc} Gaussian target}
We use the Gaussian color-source description of the McLerran--Venugopalan model \cite{McLerran:1993ni,McLerran:1993ka}, with longitudinal white noise and screened transverse covariance
\begin{align}
 d(r)&=\gamma(0)-\gamma(r)
 =\kappa\int\frac{\dd^2\ell}{(2\pi)^2}
 \frac{1-e^{i\ell r}}{(\ell^2+\Lambda^2)^2}\nonumber\\
 &=\frac{\kappa}{4\pi\Lambda^2}\,[1-\Lambda rK_1(\Lambda r)],\nonumber\\
 S_F(r)&=e^{-\CF d(r)} .
 \label{eq:target}
\end{align}
The strength $\kappa$ includes the longitudinal color-source integral. We define $Q_s$ by $\CF d(\sqrt2/Q_s)=1/2$, with $\CF=(\Nc^2-1)/(2\Nc)$. This implicit definition differs from the explicit logarithmic $Q_s$ parameter used in other MV conventions. Target Fourier transforms use $\int\dd^2b\,e^{-iqb}$ without a $(2\pi)^{-2}$ prefactor.

The Gaussian average can be evaluated exactly in its two-dimensional singlet space \cite{Fukushima:2007dy,Fukushima:2017mko,Dominguez2011,Iancu2012}. Write $a=\qb$ and
\begin{align}
 s&=d(u)+d(w),\\
 V&=d(b+au+zw)+d(b-zu-aw),\\
 W&=d(b+a(u-w))+d(b-z(u-w)),\nonumber\\
 \delta&=V-W .
\end{align}
Then
\begin{align}
 M_{11}&=-\CF s,\qquad M_{12}=M_{21}=\frac{\sqrt{\Nc^2-1}}{2\Nc}\delta,
 \nonumber\\
 M_{22}&=\frac{s}{2\Nc}-\frac{V}{\Nc}-\frac{\Nc^2-2}{2\Nc}W,\nonumber\\
 Q&=(1,\sqrt{\Nc^2-1})e^{\mathsf M}(1,0)^T .
 \label{eq:color}
\end{align}
This form follows by projecting $\sum_{i<j}d_{ij}T_i\cdot T_j$ onto the singlets of $F\otimes\bar F\otimes F\otimes\bar F$. Direct color-generator projection, coincidence limits and the vanishing of the kernel when either dipole size is zero provide separate checks. The reported benchmarks retain the full finite-$N_c$ color matrix.

\section{Expansion through fourth order in dipole size}
\subsection{Target independence at leading power}
Let $h_i=\partial_i\Phi(P)$. The leading small-dipole tensor is
\begin{equation}
 G_{ij}(b)=\CF\,f(\Nc d(b))\,d_{ij}(b),\qquad
 f(D)=\int_0^1\dd\tau\,e^{-D\tau},
\end{equation}
with Fourier decomposition
\begin{align}
 G_{ij}(q)&=\frac{G_0(q)}2\delta_{ij}
 +G_2(q)\left(\hat q_i\hat q_j-\frac{\delta_{ij}}2\right),\nonumber\\
 R_2(\phi)&=G_{ij}(q)\avg{h_i h_j^\dagger}.
 \label{eq:lp}
\end{align}
The WW tensor and its linear polarization have been studied in the MV model and under small-$x$ evolution \cite{Metz:2011wb,Dominguez:2011br,Dumitru:2015gaa,Marquet:2016cgx}. Full angular averaging removes its traceless part, so that
$\bR_2=G_0\sum_i\avg{h_i h_i^\dagger}/2$. The normalized state is independent of the target. This argument requires rotational symmetry, complete azimuthal acceptance, fixed other kinematics and the prescribed spin identification; Gaussian closure is not necessary at this order.

For a spin-summed check, put $s_z=z^2+\qb^2$. Directly tracing the spin matrix gives
\begin{align}
 \Tr R_{2,T}&=\frac{2z\qb}{D_P^4}
 \Big\{G_0[s_z(P^4+\eps^4)+2m^2P^2]\nonumber\\
 &\qquad+2G_2P^2[m^2-s_z\eps^2]\cos2\phi\Big\},\\
 \Tr R_{2,L}&=\frac{16(z\qb)^3Q^2P^2}{D_P^4}
 (G_0+G_2\cos2\phi).
 \label{eq:lptrace}
\end{align}
The full transverse spin matrix also contains a term with zero spin trace proportional to $G_2\sin2\phi$. It vanishes in the spin sum and in Eq.~\eqref{eq:average}, but must be retained before either operation.

\subsection{Expansion of the Gaussian kernel}
Define $t=1-2z$, $c_3=1-3z\qb$ and the entire functions
\begin{align}
 g(D)&=\int_0^1(1-\tau)e^{-D\tau}\dd\tau,\nonumber\\
 h(D)&=\int_0^1\tau e^{-D\tau}\dd\tau,\quad
 j(D)=\int_0^1\tau^2e^{-D\tau}\dd\tau.
\end{align}
Here $D=\Nc d(b)$.
Derivatives of $d$ below are evaluated at nonzero $b$. The geometric combinations are
\begin{align}
 \delta_2&=d_{ij}u_iw_j,\qquad
 \delta_3=\frac t2d_{ijk}u_iw_j(u-w)_k,\\
 \delta_4&=\frac{c_3}{6}
 (d_{ijkl}u_iu_ju_kw_l+d_{ijkl}u_iw_jw_kw_l)\nonumber\\
 &\quad-\frac{t^2}{4}d_{ijkl}u_iu_jw_kw_l,\\
 W_1&=t\,d_i(u-w)_i,\nonumber\\
 W_2&=\frac{s_z}{2}d_{ij}(u-w)_i(u-w)_j.
\end{align}
A Duhamel expansion of Eq.~\eqref{eq:color} gives
\begin{align}
 K_2&=\CF f\delta_2,\\
 K_3&=\CF\left[f\delta_3-\frac{\Nc h}{2}\delta_2W_1\right],\\
 K_4&=\CF\Bigg[f\delta_4-\frac{\Nc h}{2}(\delta_3W_1+\delta_2W_2)\nonumber\\
 &\quad+\frac{\Nc^2j}{8}\delta_2W_1^2
 +\left(\frac{g}{2\Nc}-\frac h\Nc\right)\delta_2^2\nonumber\\
 &\quad+\left(\frac h{2\Nc}-\CF g\right)\delta_2\{d(u)+d(w)\}
 \Bigg]\nonumber\\
 &\quad+\CF^2d(u)d(w).
 \label{eq:k4}
\end{align}
Self-dipoles in this formula are kept to order $u^2$ including their logarithms. The last term is independent of $b$ and contributes only at $q=0$ in a homogeneous target. All present benchmarks have $q>0$; a bin containing the origin needs this distribution restored.

At the operator level, for a smooth regulated field and a directional derivative, $A=(\partial U)U^\dagger$ gives
\begin{align}
 B_1&=A,\qquad B_2=tA'+A^2,\nonumber\\
 B_3&=c_3 A''+(2-3z)A'A+(1-3z)AA'+A^3.
 \label{eq:operators}
\end{align}
The order of the noncommuting factors must be preserved. Fourier factors $i^n(-i)^m/(n!m!)$ generate the $12+21$ terms at total degree three and all $13+31+22$ terms at degree four. Equation~\eqref{eq:k4} contains them together with the MV logarithms; retaining only $22$ would miss same-order interference.

Odd-rank target tensors give $R_3(\phi+\pi)=-R_3(\phi)$, whereas $R_4(\phi+\pi)=R_4(\phi)$. Hence $\bR_3=0$ without setting $z=1/2$. At fixed angle the normalization must instead retain
\begin{align}
 \delta\rho_3&=\frac{R_3-\rho_0t_3}{t_2},\nonumber\\
 \delta\rho_4&=\frac{R_4-\rho_0t_4}{t_2}-\frac{t_3}{t_2}\delta\rho_3,
 \label{eq:normalization}
\end{align}
where $t_n=\Tr R_n$ and $\rho_0=R_2/t_2$.
Every correction has zero trace. After averaging $R$ first, the last term disappears and $\delta\brho_4=(\bR_4-\brho_0\Tr\bR_4)/\Tr\bR_2$.

\subsection{Logarithmic terms and the fourth-order spin matrix}
The screened target is not analytic at a zero-size dipole:
\begin{equation}
 d(u)=\frac{\kappa u^2}{16\pi}
 \left[\ln\frac1{u^2\Lambda^2}+2\ln2-2\gamma_E+1\right]
 +O(u^4\ln u).
 \label{eq:selflog}
\end{equation}
Replacing it by an ordinary Taylor polynomial would omit scalar corrections that survive angular averaging. With $\ell_u=\ln[1/(u^2\Lambda^2)]$, the angular projection of Eq.~\eqref{eq:k4} has four invariants:
\begin{align}
 \langle K_4\rangle_{\theta_b}
 ={}&A(u\cdot w)(u^2+w^2)+B u^2w^2\nonumber\\
 &+C(u\cdot w)^2+L(u\cdot w)(u^2\ell_u+w^2\ell_w).
 \label{eq:invariants}
\end{align}
Appendix~\ref{app:coeff} gives the complete finite-$N_c$ radial coefficients.

The logarithmic hard factors can be transformed exactly:
\begin{align}
 F_{\log}(P)&=\int\frac{\dd^2u}{2\pi}e^{-iPu}\ell_u K_0(\eps u)\nonumber\\
 &=\frac2{D_P}\left[\ln\frac{D_P}{2\eps\Lambda}+\gamma_E\right],\label{eq:scalarlog}\\
 V^{\log}_i(P)&=\int\frac{\dd^2u}{2\pi}e^{-iPu}\ell_u(-i\partial_i)K_0(\eps u)\nonumber\\
 &=P_i\left[F_{\log}(P)-\frac{\ln(D_P/\eps^2)}{P^2}\right].
 \label{eq:logs}
\end{align}
Writing $\Phi(k)=(c+b_i k_i)/(k^2+\eps^2)$, we use
$\Phi_{\log}=cF_{\log}+b_iV_i^{\log}$ and its derivatives. The second term in $V_i^{\log}$ cannot be obtained by multiplying the ordinary wave function by a common logarithm.

The ordinary function $(\nabla^4d)_{b>0}$ omits the contact term $-\kappa\delta^{(2)}(b)$. Its complete transform is
\begin{equation}
 \widetilde{\nabla^4d}(q)=-\frac{\kappa q^4}{(q^2+\Lambda^2)^2}.
 \label{eq:contact}
\end{equation}
We write its coefficient $f=1+(f-1)$, transform the linear term analytically, and integrate the locally integrable nonlinear remainder. The dilute limit independently checks this prescription.

Let $a_4,b_4,c_4,\ell_4$ denote the Fourier--Bessel integrals of $(A,B,C,L)$ including Eq.~\eqref{eq:contact}. Define $h_{i\Delta}=\partial_i\nabla_P^2\Phi$ and its logarithmic counterpart. The fourth-order contribution to the averaged spin matrix is
\begin{align}
 H_1&=-\sum_i\avg{h_{i\Delta}h_i^\dagger+h_i h_{i\Delta}^\dagger},\\
 H_2&=\avg{(\nabla_P^2\Phi)(\nabla_P^2\Phi)^\dagger},\quad
 H_3=\sum_{ij}\avg{h_{ij}h_{ij}^\dagger},\\
 H_{\log}&=-\sum_i\avg{h^{\log}_{i\Delta}h_i^\dagger+
 h_i(h^{\log}_{i\Delta})^\dagger},\\
 \bR_4&=a_4H_1+b_4H_2+c_4H_3+\ell_4H_{\log}.
 \label{eq:r4}
\end{align}
The contraction retains all spin components. Fixed-angle matrices use the harmonics $\pm1,\pm3$ at cubic order and $0,\pm2,\pm4$ at quartic order, with the same logarithmic and distributional treatment.

\section{Kinematic corrections and additional current insertions}\label{sec:currents}
Kinematic corrections and terms with additional current insertions can be separated before summing over the final spins. In the gauge convention of the transverse Wilson link, let $A_i=(\partial_iU)U^\dagger$. Along its straight path,
\begin{equation}
 U(v+\qb u)U^\dagger(v-zu)
 =\mathcal P\exp\left[\int_{-z}^{\qb}\dd\alpha\,u_iA_i(v+\alpha u)\right].
\end{equation}
The one-current amplitude is therefore
\begin{align}
 \mathcal A^{[1]}_a&=iA_i(q)\mathcal H_{i,a}(P,q),\nonumber\\
 \mathcal H_i&=\int_{-z}^{\qb}\dd\alpha\,\partial_i\Phi(P-\alpha q),\nonumber\\
 R^{[1]}_{ab}&=G_{ij}(q)\avg{\mathcal H_{i,a}\mathcal H_{j,b}^*}.
 \label{eq:itmd}
\end{align}
The path identity is independent of mass and spin. It implements the established amplitude-level construction \cite{Altinoluk2019,Altinoluk:2019wyu,Boussarie:2020vzf,Boussarie2021}, here with the finite-mass spin vector retained explicitly and checked against the massive hard factors of Ref.~\cite{Altinoluk2021}. Finite-mass DIS WW hard tensors also appear in Ref.~\cite{Marquet:2017xwy}. Both WW tensor components must be kept; replacing $G_{ij}$ by a purely $\hat q_i\hat q_j$ tensor is not the full massive ITMD expression.

Expanding the hard vector gives
\begin{equation}
 \mathcal H_i=h_i-\frac t2q_jh_{ij}+\frac{c_3}{6}q_jq_kh_{ijk}+\cdots .
 \label{eq:itmdexp}
\end{equation}
Its squared amplitude contains all one-current cubic and quartic terms. In the dilute limit $G_{ij}\propto q_iq_j$, the exact identity
\begin{equation}
 q_i\mathcal H_i=\Phi(P+zq)-\Phi(P-\qb q)
 \label{eq:dilute}
\end{equation}
reproduces the coherent single-field amplitude, including its interference. At fixed $|q|$, the dilute covariance is a common factor and cancels from the normalized state even away from the correlation limit.

For a separate spin-summed test, define
\begin{align}
 J_{0i}&=-\int_{-z}^{\qb}\dd\alpha\,\partial_i(1/D_\alpha),\nonumber\\
 J_{1,ij}&=\int_{-z}^{\qb}\dd\alpha\,\partial_i(k_j/D_\alpha),
\end{align}
where $k=P-\alpha q$ and $D_\alpha=k^2+\eps^2$. Direct tracing gives
\begin{align}
 \Tr R_T^{[1]}&=2z\qb\,G_{ij}
 [m^2J_{0i}J_{0j}+s_zJ_{1,ik}J_{1,jk}],\\
 \Tr R_L^{[1]}&=8(z\qb)^3Q^2G_{ij}J_{0i}J_{0j}.
\end{align}
These agree with independent transcriptions of the finite-mass projections in Ref.~\cite{Altinoluk2021}, including nonmaximal gluon linear polarization, and with the geometric hard tensors of Ref.~\cite{Boussarie2021}.

The rest of the path expansion contains at least two current insertions. Their squared amplitudes and interference with the one-current amplitude define the genuine higher-twist sector in this scheme. At quartic order we subtract the one-current correction using the same leading normalization. The remainder is a correction to the spin state rather than a separately positive density matrix. The decomposition is defined within the stated approximation and current-insertion scheme. The one-current approximation retains the nonlinear target physics already present in the WW correlator. Gaussian closure fixes the additional correlators through the same $d(r)$, with no independent higher-point degrees of freedom.

\section{Limiting cases}
\subsection{Longitudinal photons}
Longitudinal maximal entanglement has been established in photon--gluon fusion \cite{Qi2026}, diffraction \cite{FucillaDiffractive2026}, and the inclusive correlation limit \cite{Fucilla2026}. The diffractive amplitude in Ref.~\cite{FucillaDiffractive2026} explicitly separates spin from target dependence. For the inclusive target trace beyond that limit, Eq.~\eqref{eq:wavefunctionL} factorizes the longitudinal spin vector $B_a=(\bar u\gamma^+v)_a$ from every internal momentum. At fixed external kinematics, for any spin-independent eikonal target,
\begin{align}
 \mathcal A^L_a&=B_a\,\mathcal I_A,\qquad R^L=F_A BB^\dagger,\nonumber\\
 \rho^L&=\frac{BB^\dagger}{B^\dagger B}\quad(F_A>0).
 \label{eq:longitudinal}
\end{align}
The transported $B$ does not depend on $\phi$, so full angular averaging preserves the result. Its two Schmidt coefficients are equal and $\cN=1/2$. Equation~\eqref{eq:longitudinal} holds to all orders in eikonal scattering under these assumptions. It does not apply to a transverse/longitudinal mixture, broad kinematic bins, a different angle-dependent spin identification, or unresolved radiation. The next-to-soft radiative mechanism of Ref.~\cite{Agrawal2026} lies beyond this tree-level treatment.

\subsection{Mass dependence of the transverse correction}
For $\Phi^\lambda=(c^\lambda+b_i^\lambda k_i)/(k^2+\eps^2)$ define
$M_0=\avg{cc^\dagger}$, $M_1=\avg{b_xc^\dagger+cb_x^\dagger}$ and
$M_b=\avg{b_xb_x^\dagger}$. Light-front spin algebra gives
\begin{align}
 \avg{b_yb_y^\dagger}&=M_b,\qquad
 \avg{b_xb_y^\dagger+b_yb_x^\dagger}=0,\nonumber\\
 \Tr M_0&=2z\qb m^2,\qquad \Tr M_b=2z\qb s_z,\nonumber\\
 \Tr M_1&=0,\qquad \fnorm{M_1}=2z\qb m\sqrt{s_z}.
 \label{eq:massidentities}
\end{align}
For the real reflection-symmetric Gaussian kernel after complete angular averaging, the $y$-directed mass interference vanishes. In the integrable limit $m\to0$ at fixed $Q^2>0$ and $0<z<1$, only $M_b$ remains: the averaged state is $M_b/\Tr M_b$, with $\cN=z\qb/s_z$. The massless limit provides a consistency check on the mass-dependent correction.

At fixed $m,Q^2,z,q$ and target parameters, the leading matrix has
$\bR_2=G_0[M_b/P^4+M_1/P^5+O(P^{-6})]$. The leading $M_b$ terms of $\bR_4$ cancel in its normalization. Expanding the exact hard factors in Eq.~\eqref{eq:r4} yields
\begin{equation}
 \delta\brho_4=
 \frac{-16a_4-32\ell_4\left[
 \ln\!\frac{P^2}{2\eps\Lambda}+\gamma_E-\frac94\right]}
 {G_0\,\Tr M_b\,P^3}\,M_1+O(P^{-4}\ln P).
 \label{eq:largep}
\end{equation}
Thus its leading norm is proportional to $m/P^3$ times a logarithmic coefficient; a rate-level relative $1/P^2$ correction cannot be assigned directly to the normalized state. The terms proportional to $b_4$ and $c_4$ cancel from this leading mass-interference coefficient. The limit holds with $Q^2$ fixed, not with $Q/P$ fixed.

Two separately written calculations reproduce Eq.~\eqref{eq:largep}: Cartesian differentiation of the scalar numerator basis and a radial expansion in $P^2$. Before normalization, the coefficients of $M_b/P^6$ in $(H_1,H_2,H_3,H_{\log})$ are $(0,0,16,32)$. The corresponding coefficients of $M_1/P^7$ are $(-16,0,16,104-32[\ln(P^2/(2\eps\Lambda))+\gamma_E])$. Subtracting the leading-state trace contribution gives $-16,0,0$ and the logarithmic coefficient in Eq.~\eqref{eq:largep}. Direct Dirac-spinor checks of Eq.~\eqref{eq:massidentities} and coordinate-space evaluations of Eqs.~\eqref{eq:scalarlog},\eqref{eq:logs} test the inputs to these algebraic calculations. Deterministic evaluations of the quartic term at $q/m=0.2$ and the two target strengths give asymptotic-to-exact norm ratios about $0.9978$ at $P/m=32$ and $0.99987$ at $128$. The relative matrix remainder approximately halves at each doubling from 32 to 128. These checks establish the asymptotic behavior of the calculated quartic term; full-kernel validation at these larger hard scales remains to be performed. At the $P/m=2$ reference point, the asymptotic expression is not accurate.

\section{Numerical method and validation}
\subsection{Deterministic expansion and full-kernel reference}
For the original benchmark scan, dimensional quantities are expressed in units of the fixed quark mass $m=m_0$. The subsequent mass scan keeps this reference unit fixed, as specified in Sec.~\ref{sec:robustness}. The target parameters are $\Nc=3$, $\Lambda/m=0.05$ and $Q_s/m=0.25,0.50$. We fix $z=0.35$, $Q^2/m^2=4$ and use $q/m=0.1,0.2,0.4$. The expansion data cover $P/m=2,4,8$ for both photon channels. At $P/m=2$, $q/\sqrt{D_P}=0.0411$--$0.1645$ and $Q_s/\sqrt{D_P}=0.1028,0.2057$. These ratios are diagnostics, not sufficient bounds on the MV high-momentum tail or the truncation error.

One-dimensional Fourier--Bessel quadrature evaluates Eq.~\eqref{eq:r4}. Tightening tolerances and moving the endpoint from $\Lambda b_{\max}=35$ to 45 changes the normalized quartic results by less than $7.3\times10^{-14}$ in Frobenius norm. Direct coordinate-space integration checks 16 logarithm-weighted hard moments with maximum absolute residual $9.3\times10^{-15}$. The complete fixed-angle implementation uses an exact finite Fourier projection of the allowed angular polynomials and radial vector quadrature. The calculation contains 144 fixed-angle records for 36 choices of kinematics and photon polarization. The 36 angular averages agree with the separate scalar projection to $9.6\times10^{-15}$. Tightening the direction grid, radial tolerance and tail changes normalized matrices by at most $1.1\times10^{-9}$. Fixed-angle full-kernel truncation checks have not been performed.

Recent numerical treatments also evaluate DIS kernels beyond the correlation limit \cite{Duan:2026yrk,Duan:2026jrk}; their demonstrated spin-summed, massless large-$N_c$ implementation differs from the finite-mass calculation with explicit spin indices here. For the full kernel at $q>0$, only the connected combination $Q-S_F(u)S_F(w)$ remains after Fourier transformation in $b$. We integrate the common transverse rotation analytically, leaving five variables: $|b|,|u|,|w|$ and two relative angles. The full matrix is represented by six basis matrices in spin space; the Bessel weights were checked against a separate 512-direction integral. We subtract the exact dilute contribution, the nonlinear leading term and, optionally, the nonlinear quartic term as control variates. Their analytically or deterministically integrated values are restored. The unexpanded two-channel exponential is always used in the remainder. Appendix~\ref{app:integrator} specifies the measure, basis, importance maps, and control restoration.

The original scan uses eight independent scrambled Sobol sequences per reference matrix; the precision study uses 16 or 32. Samples are mapped to strictly interior points of a 52-bit dyadic cell grid; every chunk is checked for nonfinite integrands. Each of the eight parameter points has a coarse and a fine run with the quartic control and a coarse run with the leading-power control. Per scramble, the sample counts are $2^{20},2^{22},2^{20}$ at $P/m=2$ and $2^{21},2^{24},2^{21}$ at $P/m=4$.

\subsection{Correlated uncertainty and comparison criteria}
Let $R_r$ be the unnormalized matrix from scramble $r$, $n$ independent scrambles, and $\widehat R=n^{-1}\sum_rR_r$. We normalize the mean, and propagate its trace covariance using the influence matrices
\begin{align}
 \widehat\rho&=\frac{\widehat R}{\Tr\widehat R},\qquad
 X_r=\frac{R_r-\widehat R-\widehat\rho(\Tr R_r-\Tr\widehat R)}
 {\Tr\widehat R},\\
 \sigma_{\mathrm F}^2&=\frac1{n(n-1)}\sum_r\fnorm{X_r-\bar X}^2 .
 \label{eq:error}
\end{align}
For target contrasts we pair the same scramble seeds and use the difference of the influence matrices; the importance mappings depend on the target. Treating the two estimates as independent would discard this covariance. For negativity we propagate through the negative-eigenspace projector of $\rho^{T_{\bar Q}}$ and compare with a leave-one-scramble-out jackknife. These agree at the reported points, which are away from the boundary where the partial transpose first develops a negative eigenvalue.

Here $\brho_{\rm quartic}=\brho_0+\delta\brho_4$ denotes the consistently expanded normalized state, rather than $(\bR_2+\bR_4)/\Tr(\bR_2+\bR_4)$. With $E=\fnorm{\delta\brho_4}$, the numerical requirements fixed before the reference calculations are
\begin{equation}
 \frac{4\sigma_{\mathrm F}}{E}\le0.15,\qquad
 \frac{\fnorm{\widehat\rho-\brho_{\rm quartic}}+4\sigma_{\mathrm F}}{E}\le0.25 .
 \label{eq:gates}
\end{equation}
The sample-increase and control-choice differences must each be no larger than four paired standard errors. We inspect positivity on the original matrix, retaining its eigenvalues and components without clipping or symmetry projection. The factor four defines an empirical diagnostic based on a finite set of scrambles; it is not a calibrated simultaneous confidence level or an estimate of all theory uncertainties.

All eight points satisfy Eq.~\eqref{eq:gates} and the paired checks. Precision fractions range from $0.650\%$ to $14.012\%$, and the truncation diagnostic from $4.518\%$ to $19.097\%$. The smallest central density eigenvalue exceeds $0.0298$. In the original eight-point dataset the $P/m=4$ points are close to the precision threshold; the separate refinement in Sec.~\ref{sec:robustness} uses new scrambles. In the original scan, the coarse control-choice uncertainties are $114\%$ and $123\%$ of $E$, so this comparison has limited resolving power. The controls also share parts of the same physical kernel. Checks using other representations include the color-generator construction, hard-coordinate integrals, dilute amplitude, literature hard tensors, and tensor-versus-scalar quartic projection. Numerical uncertainties for the additional precision study are listed in Appendix~\ref{app:uncertainties}.

\section{Results and physical interpretation}
\begin{figure*}[t]
\centering\includegraphics[width=\textwidth]{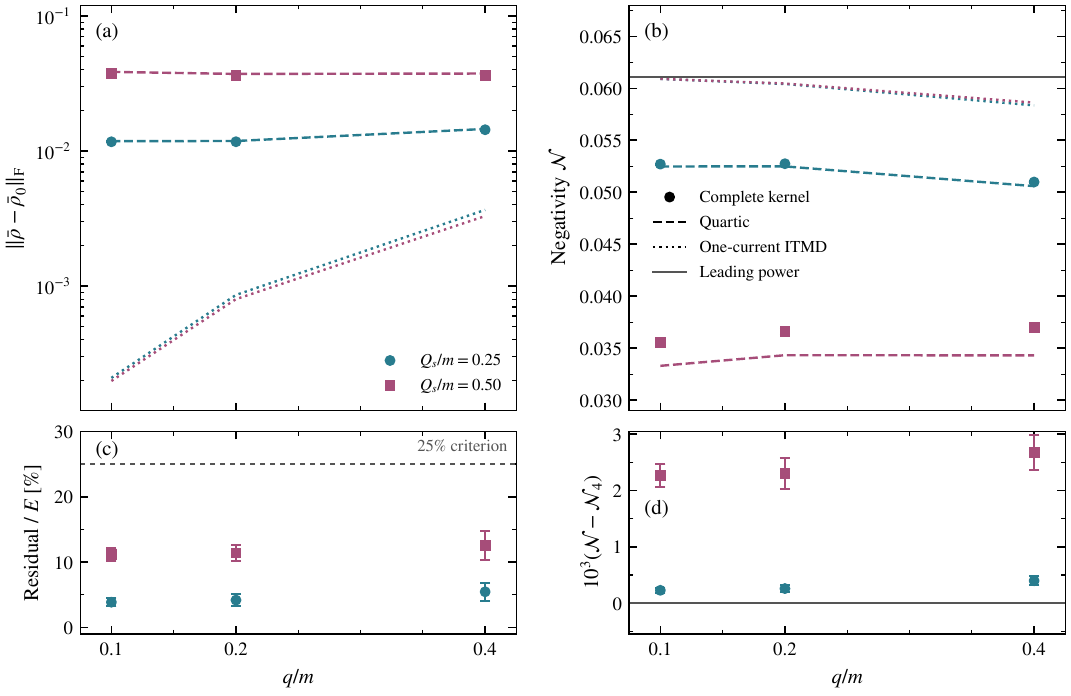}
\caption{Transverse-photon correction to the spin state and negativity at $P/m=2$, with the remaining parameters of Table~\ref{tab:points}. Top panels compare the original eight-scramble full Gaussian estimates with the quartic expansion and one-current ITMD result. Bottom panels show the difference between the full and fourth-order spin matrices divided by $E$, and the negativity residual multiplied by $10^3$. In panel (c), the upper endpoint of each bar is the truncation diagnostic of Eq.~\eqref{eq:gates}, to be compared with $25\%$. Colors and marker shapes distinguish target strengths. Bars denote four empirical scramble standard errors, not confidence intervals. Dashed and dotted segments join discrete approximation values; the full-kernel points are unconnected.}
\label{fig:main}
\end{figure*}

\begin{figure*}[t]
\centering\includegraphics[width=\textwidth]{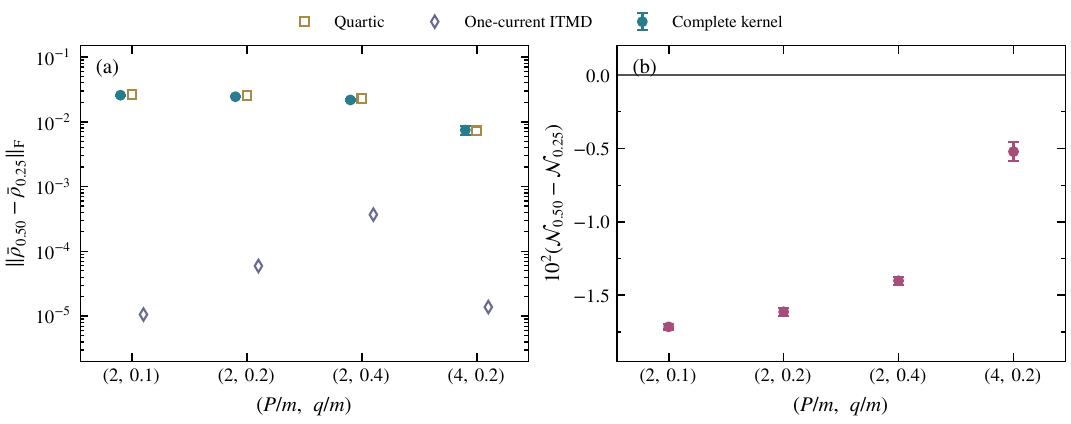}
\caption{Transverse-photon target contrasts in the original eight-scramble scan at the four matched kinematic pairs, with $z=0.35$, $Q^2/m^2=4$, $\Lambda/m=0.05$ and $N_c=3$. The left panel compares the complete, quartic and one-current matrix distances on a logarithmic scale; marker offsets separate the methods at each discrete pair. The right panel shows the complete negativity difference multiplied by $10^2$. Error bars include the same-seed target covariance and denote four empirical standard errors. All four lower matrix-distance diagnostics are positive.}
\label{fig:contrast}
\end{figure*}

\begin{figure*}[t]
\centering\includegraphics[width=0.86\textwidth]{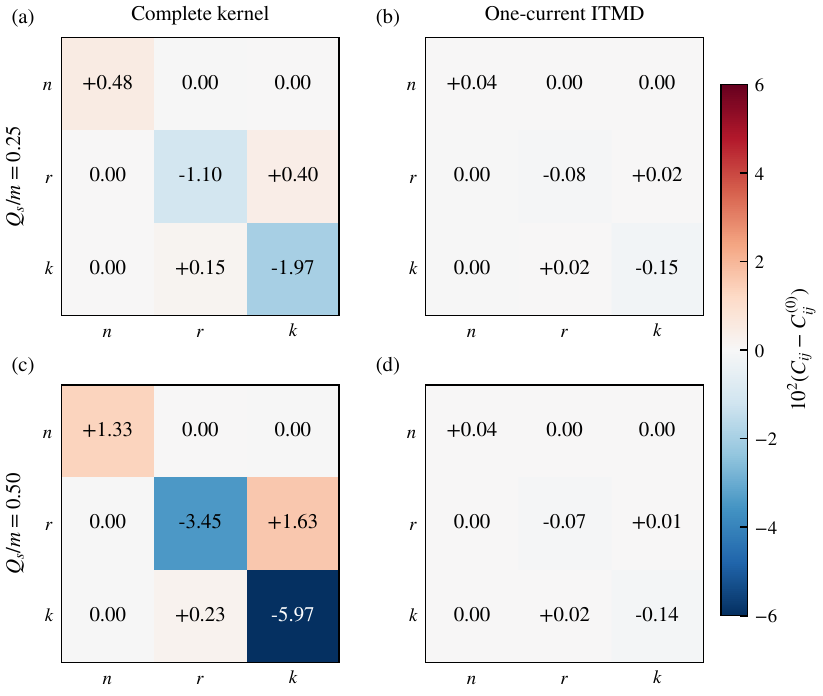}
\caption{Transverse-photon changes in the correlation tensor $C_{ij}$ from leading power at $P/m=2,q/m=0.2$, $z=0.35$, $Q^2/m^2=4$, $\Lambda/m=0.05$ and $N_c=3$. Rows correspond to $Q_s/m=0.25,0.50$ and columns to full Gaussian and one-current ITMD results. Colors and labels show $10^2(C_{ij}-C_{ij}^{(0)})$ on one common symmetric scale. The complete values are unprojected central estimates from the original eight-scramble scan. Rounded zeros can contain sampling noise; the displayed digits do not assert equal component precision. The colors and printed values refer to central estimates.}
\label{fig:tensor}
\end{figure*}

Tables~\ref{tab:points} and \ref{tab:contrasts}, and Figs.~\ref{fig:main}--\ref{fig:tensor}, show the original scan. Its central and $P/m=4$ pairs are independently repeated with higher precision in Sec.~\ref{sec:robustness}; both estimates are shown to distinguish the original calculation from the precision study. At $P/m=2$, the leading-power transverse negativity is $0.06106018$, independent of the target. Every full-kernel benchmark in that slice gives a smaller negativity. The quartic correction reproduces its target dependence, with the residual quantified separately from the sampling uncertainty by Eq.~\eqref{eq:gates}.

\begin{table*}[t]
\centering\small
\begin{tabular}{rrrrrrr}
\toprule
$P/m$ & $Q_s/m$ & $q/m$ & $T$ & $\|\bar\rho-\bar\rho_{\rm LP}\|_{\rm F}$ & $\mathcal N$ & $\sigma_{\mathcal N}$ \\
\midrule
2 & 0.25 & 0.1 & 0.0460635 & 0.0117066 & 0.0527105 & 0.0000107 \\
2 & 0.25 & 0.2 & 0.0279074 & 0.0117170 & 0.0527485 & 0.0000157 \\
2 & 0.25 & 0.4 & 0.0108585 & 0.0143774 & 0.0509877 & 0.0000202 \\
2 & 0.50 & 0.1 & 0.0729971 & 0.0374524 & 0.0355637 & 0.0000499 \\
2 & 0.50 & 0.2 & 0.0530265 & 0.0360795 & 0.0366246 & 0.0000689 \\
2 & 0.50 & 0.4 & 0.0289043 & 0.0360949 & 0.0369770 & 0.0000769 \\
4 & 0.25 & 0.2 & 0.0031092 & 0.0032964 & 0.2839633 & 0.0000764 \\
4 & 0.50 & 0.2 & 0.0060937 & 0.0107678 & 0.2787477 & 0.0002207 \\
\bottomrule
\end{tabular}

\caption{Transverse-photon full Gaussian reference from the original eight-scramble scan at $z=0.35$, $Q^2/m^2=4$, $\Lambda/m=0.05$ and $\Nc=3$. Momenta are in units of $m$. $T=\Tr\bR$ is the reduced rate in the wave-function convention of Eq.~\eqref{eq:wavefunction}; it omits common charge, flux and phase-space factors. The last column is one empirical standard error on negativity. The Frobenius distance is to the correctly averaged leading-power state.}
\label{tab:points}
\end{table*}

The target comparison is made at identical kinematics. At $P/m=2,q/m=0.2$, the two complete states differ by $\fnorm{\brho_{0.50}-\brho_{0.25}}=0.02440707$, with paired $\sigma_{\mathrm F}=0.00011185$. Its empirical lower diagnostic $D-4\sigma_{\mathrm F}=0.02395968$ is positive. Negativity changes from $0.05274855$ to $0.03662460$; the difference is $-0.01612395$ with empirical four-standard-error interval $[-0.01638236,-0.01586554]$. The other target pairs give the same qualitative conclusion (Table~\ref{tab:contrasts}, Fig.~\ref{fig:contrast}).

\begin{table*}[t]
\centering\small
\begin{tabular}{rrrrrr}
\toprule
$P/m$ & $q/m$ & $D$ & $D-4\sigma_{\rm F}$ & $\Delta\mathcal N\ \pm4\sigma$ & $D_{\rm one}$ \\
\midrule
2 & 0.1 & 0.0257922 & 0.0254479 & $-0.0171468\pm0.0001840$ & 0.0000106 \\
2 & 0.2 & 0.0244071 & 0.0239596 & $-0.0161239\pm0.0002585$ & 0.0000595 \\
2 & 0.4 & 0.0217786 & 0.0211088 & $-0.0140107\pm0.0002618$ & 0.0003692 \\
4 & 0.2 & 0.0074722 & 0.0062128 & $-0.0052155\pm0.0006615$ & 0.0000138 \\
\bottomrule
\end{tabular}

\caption{Transverse-photon contrasts from the original eight-scramble scan between $Q_s/m=0.50$ and $0.25$. $D$ is the Frobenius distance between the full spin states; the displayed lower value uses four paired standard errors and is rounded downward. The $\Delta\cN$ interval uses four empirical standard errors and is rounded outward. $D_{\rm one}$ is the corresponding one-current ITMD distance. The error convention has the limitations stated below Eq.~\eqref{eq:gates}.}
\label{tab:contrasts}
\end{table*}

The one-current contrast at the central benchmark is $5.9452\times10^{-5}$, much smaller than the complete value $0.0244071$. Kinematic corrections give a small nonzero target contrast through the ratio $G_2/G_0$. The dominant target contrast at these points resides in the terms with additional current insertions defined in Sec.~\ref{sec:currents}. Figure~\ref{fig:tensor} displays the full spin-correlation change, including its off-diagonal components, in the specified transported basis.

At $q/m=0.2$, increasing $P/m$ from 2 to 4 reduces the complete distance from leading power from $0.01172$ to $0.00330$ for the weaker target and from $0.03608$ to $0.01077$ for the stronger target. This supports approach toward the leading-power state on the sampled slice. Two full-kernel hard scales cannot determine an asymptotic exponent; Eq.~\eqref{eq:largep} and its deterministic quartic controls provide a separate analytic statement. There is no full-kernel validation at $P/m=8$ or at fixed azimuth.

The two targets already have distinct reduced unpolarized rates: at the central benchmark they are $0.02790740$ and $0.05302652$. Section~\ref{sec:information} quantifies the information added by spin with a common conditional likelihood after profiling the total normalization. Gaussian closure fixes the finite-$N_c$ quadrupole throughout.

\section{Infrared and kinematic robustness}\label{sec:robustness}
The precision and robustness study uses statistically independent scrambles of the same physical kernel and fixes a reference unit $m_0$, equal to the original
benchmark mass. All momenta below are in units of $m_0$ and virtualities in
$m_0^2$. Varying $m$ therefore changes the physical mass at fixed values of the other dimensional inputs. At fixed $q=0.2$ and $Q_s=0.25,0.50$, we change
$\Lambda=0.025,0.05,0.10$, $z=0.25,0.35,0.45$, $Q^2=1,4,9$ and
$m=0.75,1,1.25$ individually and include two combined changes. Except for
the $P=4$ pair, $P=2$. The implicit definition of $Q_s$ is held fixed;
the source strength $\kappa$ is recalculated and saved at each point.
Changing $\Lambda$ probes the screening prescription within the target model; it is not a regulator-removal limit.

The refined estimates use statistically independent 52-bit scrambles, with seeds disjoint
from the original eight. Each sample evaluates the complete kernel once
and accumulates both LP and quartic controls. Their separately integrated
control terms are restored before normalization. At $P=2$, 16 scrambles each contain $2^{22}$ fine points with a nested $2^{20}$ prefix. At $P=4$, 32 scrambles each contain $2^{26}$ fine points with a nested $2^{25}$ prefix. The target contrasts retain
same-seed covariance; comparisons with the original estimates combine independent
variances. The controls share the underlying physical kernel.

The precision requirements tighten to $4\sigma_{\mathrm F}/E\le0.10$
at $P=2$ and $0.05$ at $P=4$, while the truncation bound remains $0.25$.
Both nested-sample and control-choice distances must be within four
paired standard errors. In addition, the uncertainty of the control
difference itself must obey $4\sigma_{\rm ctrl}/E\le0.25$ at $P=2$ and
$0.15$ at $P=4$. Positivity with a four-standard-error margin,
Hermiticity, normalization and deterministic refinement are checked on
unprojected matrices. These remain empirical diagnostics, without a
simultaneous-coverage calibration.

\begin{table*}[t]\centering\small
\begin{tabular}{lrrrrrrrr}
\toprule
Case & $\Lambda$ & $z$ & $Q^2$ & $m$ & $D$ & $D-4\sigma$ & $\Delta\cN$ & Gate \\
\midrule
Base & 0.05 & 0.35 & 4 & 1 & 0.02447 & 0.02414 & -0.01622 & yes \\
IR low & 0.025 & 0.35 & 4 & 1 & 0.02498 & 0.02446 & -0.01676 & yes \\
IR high & 0.1 & 0.35 & 4 & 1 & 0.02363 & 0.02340 & -0.01523 & yes \\
$z$ low & 0.05 & 0.25 & 4 & 1 & 0.02437 & 0.02397 & -0.01059 & yes \\
$z$ high & 0.05 & 0.45 & 4 & 1 & 0.02451 & 0.02422 & -0.01630 & yes \\
$Q^2$ low & 0.05 & 0.35 & 1 & 1 & 0.02997 & 0.02889 & -0.02312 & yes \\
$Q^2$ high & 0.05 & 0.35 & 9 & 1 & 0.01979 & 0.01971 & -0.00231 & yes \\
$m$ low & 0.05 & 0.35 & 4 & 0.75 & 0.02396 & 0.02333 & -0.01734 & yes \\
$m$ high & 0.05 & 0.35 & 4 & 1.25 & 0.02249 & 0.02234 & 0.00000 & yes \\
Combined 1 & 0.025 & 0.25 & 9 & 1 & 0.02097 & 0.02074 & -0.00583 & yes \\
Combined 2 & 0.1 & 0.45 & 1 & 1 & 0.02949 & 0.02882 & -0.02368 & yes \\
$P=4$ & 0.05 & 0.35 & 4 & 1 & 0.00750 & 0.00714 & -0.00521 & yes \\
\bottomrule
\end{tabular}

\caption{All additional transverse-photon paired slices in fixed $m_0$ units, with $q/m_0=0.2$ and $Q_s/m_0=0.25,0.50$. $D$ is the full
matrix distance between the two targets. Displayed lower diagnostics are rounded downward. A resolved difference requires the numerical checks and $D-4\sigma>0$. The negativity difference is recorded
separately; its uncertainty is in the data and Fig.~\ref{fig:robustness}.}
\label{tab:robustness}\end{table*}

\begin{figure*}[t]\centering
\includegraphics[width=\textwidth]{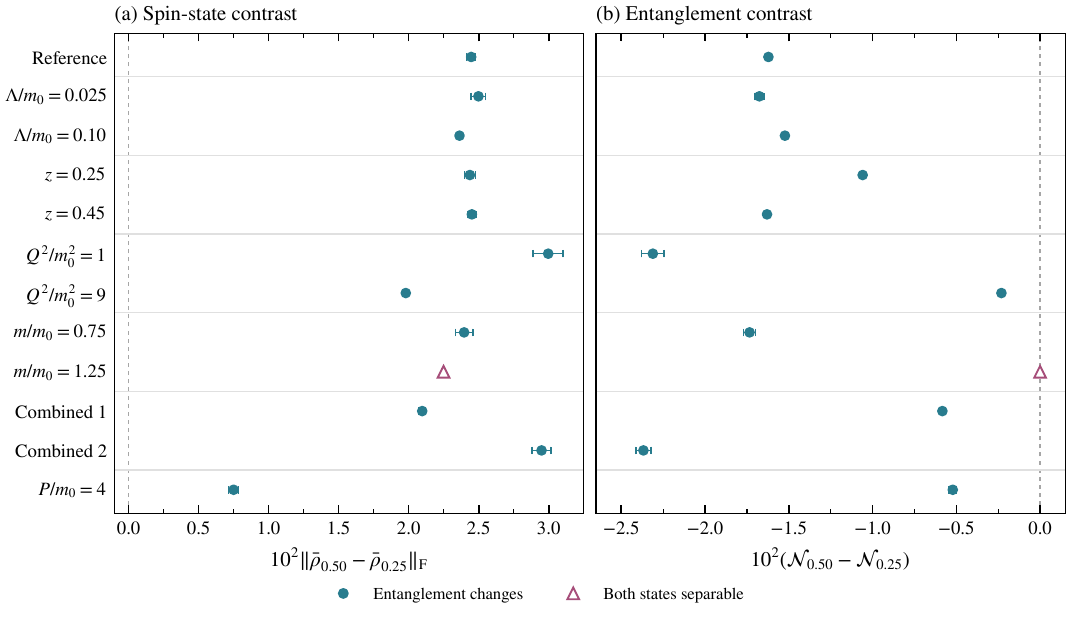}
\caption{Transverse-photon target contrasts from the refined study with error bars equal to four paired empirical standard errors. Here $q/m_0=0.2$, $Q_s/m_0=0.25,0.50$, and $N_c=3$; $P/m_0=2$ except for the labeled $P=4$ slice. The baseline has $m/m_0=1$, $z=0.35$, $Q^2/m_0^2=4$, and $\Lambda/m_0=0.05$; variations are specified in Table~\ref{tab:robustness}.
All 12 slices meet the matrix-contrast diagnostic. The open triangle identifies the larger-mass slice: both states are separable despite their resolved matrix contrast. Vertical dashed lines indicate zero contrast. The points are discrete benchmarks.}
\label{fig:robustness}\end{figure*}
All 24 computations meet the numerical requirements. Four repeat the original baseline and $P=4$ kinematics with disjoint scramble seeds, giving 28 distinct parameter points across the two studies. Their comparisons with the statistically independent original estimates are consistent. The precision fractions in the additional study range from $0.273\%$ to $3.563\%$ of $E$, and the truncation diagnostics range from $3.544\%$ to $13.431\%$. The smallest unprojected density eigenvalue exceeds $0.00406$.

\subsection{State changes and the separable pair}
All 12 paired slices have $D-4\sigma_{\mathrm F}>0$, while only 11 have a resolved negativity difference. The refined baseline gives $D=0.0244671$ with paired $\sigma_{\mathrm F}=0.0000802$ and $\Delta\cN=-0.0162190$ with a paired standard error of $0.0000387$. At $m/m_0=1.25$, however, both targets give $\cN=0$ despite $D=0.0224949$ and $D-4\sigma_{\mathrm F}=0.0223497$. Their partial transposes are positive: the smallest partial-transpose eigenvalue minus $4\sigma_{\mathrm F}$ is $0.02912$ and $0.04100$, respectively. For two qubits, positive partial transpose is equivalent to separability \cite{Peres1996,Horodecki1996}. This pair demonstrates that target-dependent spin correlations need not be target-dependent entanglement. The four-standard-error margins are sampling diagnostics, not exact bounds on all uncertainties.

Evaluating Eq.~\eqref{eq:paulidistance} on the saved matrices, the local
polarization differences vanish to numerical roundoff in all 12 pairs.
The state distance is therefore carried by $\Delta C_{ij}$ at these
points. In the transported $(n,r,k)$ axes, the baseline longitudinal
spin correlation changes by
$\Delta C_{kk}=-0.04017\pm0.00037$; for the separable $m/m_0=1.25$
pair it changes by $-0.03524\pm0.00018$.
These bars denote four paired standard errors across scrambles, obtained by projecting the
same matrix influence estimates onto $\sigma_k\otimes\sigma_k$.
They quantify individual correlations, without a simultaneous-coverage
claim or an assumption about experimental analyzing powers.

\subsection{Infrared prescription and hard-scale precision}
At fixed implicit $Q_s$, lowering $\Lambda/m_0$ from $0.05$ to $0.025$ changes the central target distance by $+2.11\%$; raising it to $0.10$ changes the distance by $-3.42\%$. The corresponding ratios are $1.0211$ and $0.9658$. The contrast thus persists with a modest amplitude change across the three tested screening prescriptions. The $z$, $Q^2$, mass, and combined variations in Table~\ref{tab:robustness} also retain a resolved state contrast. These tests cover the discrete parameter choices listed in the table.

\begin{table*}[t]\centering\small
\begin{tabular}{lrrrr}
\toprule
$Q_s$ & Old precision & New precision & Old control & New control \\
\midrule
0.25 & 13.76 & 2.99 & 113.78 & 7.94 \\
0.50 & 14.01 & 3.56 & 122.60 & 6.05 \\
\bottomrule
\end{tabular}

\caption{Transverse-photon precision and control-difference uncertainty at $P/m_0=4$, as
percentages of $E$. Target strengths are $Q_s/m_0$. The original control comparison used coarse samples;
the refined comparison uses both controls on the full fine sample.}
\label{tab:precision}\end{table*}
At $P/m_0=4$, the four-standard-error precision fraction decreases from $13.76\%,14.01\%$ in the original estimates to $2.995\%,3.562\%$ for $Q_s/m_0=0.25,0.50$. The control-difference uncertainty decreases from $113.78\%,122.60\%$ to $7.939\%,6.050\%$ of $E$, resolving the limitation of the original coarse control comparison. The refined truncation diagnostics are $4.57\%$ and $8.11\%$. These uncertainties quantify numerical sampling and quartic truncation; evolution and radiative corrections require separate calculations.

\section{Spin information conditional on the unpolarized shape}\label{sec:information}
The two targets differ in both rate and spin state. To separate these sources of discrimination, we compare their joint kinematic and spin likelihood with the likelihood of the same events after spin outcomes are discarded. Combining spin-state and cross-section information has precedents, including the trace-distance analyses of Ref.~\cite{Fabbrichesi:2025ywl}. Here we use the ordinary likelihood of specified local spin measurements. We test two fixed Gaussian hypotheses, with their target parameters held fixed throughout.

\subsection{A conditional likelihood for three narrow bins}
Let $A$ and $B$ denote $Q_s/m=0.25$ and $0.50$. We use the original full-kernel transverse matrices at $P/m=2$, $z=0.35$, $Q^2/m^2=4$, $\Lambda/m=0.05$ and $N_c=3$. The three radial bins are centered at $q_i/m=0.1,0.2,0.4$ and have a common infinitesimal width $\dd q$. All other kinematics are fixed and the relative azimuth is fully averaged. The conditional bin probabilities are
\begin{equation}
 p_{hi}=\frac{w_i T_{hi}}{\sum_j w_jT_{hj}},\qquad
 T_{hi}=\Tr\bR_h(q_i),\qquad w_i=q_i,
 \label{eq:cellprob}
\end{equation}
where $h=A,B$. The factor $q_i$ is the radial phase-space Jacobian. Equal $\dd(q^2)$ bins instead have $w_i=1$. This construction uses point values in the narrow-bin limit; finite experimental bins would require integration over their widths.

\begin{figure*}[t]\centering
\includegraphics[width=\textwidth]{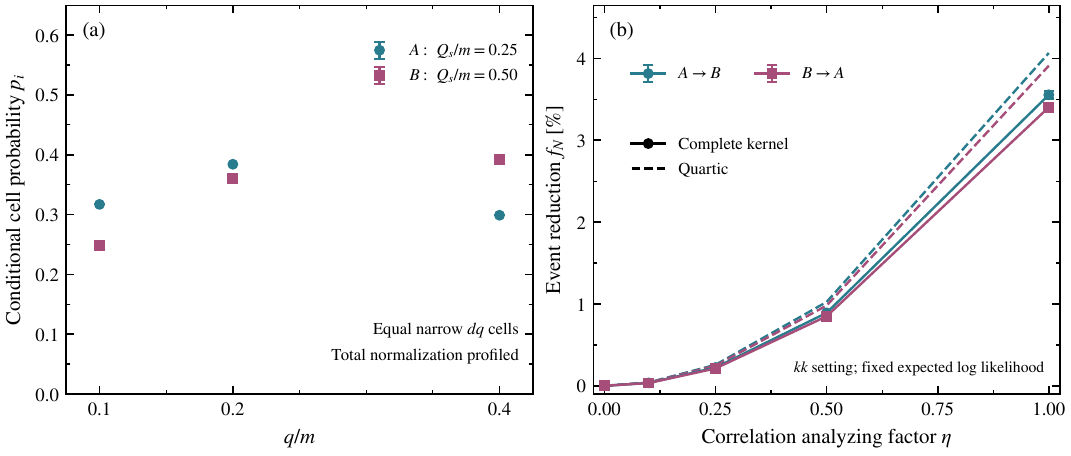}
\caption{Transverse-photon unpolarized information retained after profiling the overall normalization, and the spin increment from the same selected events. (a) The three bin probabilities of Eq.~\eqref{eq:cellprob}; the two shapes are not degenerate. (b) Fractional event reduction at fixed expected log likelihood ratio for the $kk$ protocol. Solid points use the complete kernel; dashed segments use the quartic approximation. Segments connect the specified analyzing factors. Bars are four paired numerical standard errors and are smaller than some markers. Both panels use the kinematics of Sec.~\ref{sec:information}.}
\label{fig:information}\end{figure*}

We condition on the selected event count $N$. Equivalently, consider an extended Poisson likelihood with mean bin counts $\nu_h w_iT_{hi}$ and maximize over a positive overall normalization under each hypothesis, as in standard profile-likelihood methods \cite{Cowan2011}. The maximum is $\widehat\nu_h=N/\sum_iw_iT_{hi}$. Substitution leaves the multinomial likelihood of Eq.~\eqref{eq:cellprob}, up to hypothesis-independent factors. Each hypothesis has one normalization shared by all bins. The unpolarized shapes remain different, as shown in Fig.~\ref{fig:information}(a).

For each selected event, measure both spins along their transported $k$ axes with local positive operators
\begin{equation}
 E_s=\frac{I+s\sqrt\eta\,\sigma_k}{2},\qquad s=\pm1,
 \qquad 0\le\eta\le1.
\end{equation}
All four outcomes $(s,t)$ are retained. Using the Pauli decomposition of Sec.~\ref{sec:spin}, their probabilities in bin $i$ are
\begin{equation}
 \pi_{hi,st}=\frac{1+s\sqrt\eta\,a_{hi,k}
                 +t\sqrt\eta\,b_{hi,k}+st\eta C_{hi,kk}}4.
 \label{eq:spinprob}
\end{equation}
This explicitly specified measurement is narrower than reconstruction of a general helicity density matrix \cite{Bernal:2023jba}. The ideal measurement has $\eta=1$; smaller values model a known attenuation of spin information. The attenuation is a model input; fragmentation analyzing powers and detector efficiencies have not been calibrated for this observable. We assume that the selected events have both spin outcomes available. Selection losses require a separate yield calculation.

For $A$ as the generating hypothesis, the Kullback--Leibler divergences \cite{Kullback1951} are
\begin{align}
 D_U^{A\to B}&=\sum_i p_{Ai}\ln\frac{p_{Ai}}{p_{Bi}},\nonumber\\
 D_S^{A\to B}&=\sum_i p_{Ai}\sum_{s,t}
       \pi_{Ai,st}\ln\frac{\pi_{Ai,st}}{\pi_{Bi,st}},\nonumber\\
 D_{\rm joint}^{A\to B}&=D_U^{A\to B}+D_S^{A\to B}.
 \label{eq:kl}
\end{align}
The last equality follows from the chain rule after inserting the joint probabilities $p_{hi}\pi_{hi,st}$. Thus $D_S$ is the spin increment after retaining the same unpolarized shape constraint. Its usefulness depends on its magnitude and on the measurement. We report both directions because the divergence is not symmetric.

For $N$ independent selected events, the expected log likelihood ratio is exactly $ND$. At a fixed value of this expectation, adding the spin outcomes reduces the required event count by the fraction
\begin{equation}
 f_N=\frac{D_S}{D_U+D_S}.
 \label{eq:eventfraction}
\end{equation}
This comparison does not fix test power, a median significance or a confidence level. With a known common normalization, the unpolarized Poisson information per expected $A$ event would instead be
\begin{equation}
 D_{U,\mathrm{known}}=D_U+r-1-\ln r,\qquad
 r=\frac{\sum_iw_iT_{Bi}}{\sum_iw_iT_{Ai}}.
 \label{eq:knownnorm}
\end{equation}
The conditional spin increment remains $D_S$.

\subsection{Magnitude, analyzing factor and bin dependence}
For the three equal-$\dd q$ bins, $D_U$ is $0.0218252$ in the $A\to B$ direction and $0.0222588$ in the reverse direction. At $\eta=1$, the complete kernel gives $D_S=0.0008046$ and $0.0007843$, respectively. The increments are $3.69\%$ and $3.52\%$ of the unpolarized information. Equivalently, $f_N=3.56\%$ and $3.40\%$ at fixed expected log likelihood ratio. The numerical sampling uncertainty is small compared with these increments (Table~\ref{tab:information}).

\begin{table*}[t]\centering\small
\begin{tabular}{ccrrr}
\toprule
$\eta$ & Direction & $10^4D_S$ & $100D_S/D_U$ & $100f_N$ \\
\midrule
0.25 & $A\to B$ & $0.488\pm0.007$ & $0.223\pm0.004$ & $0.223\pm0.003$ \\
0.25 & $B\to A$ & $0.473\pm0.008$ & $0.213\pm0.004$ & $0.212\pm0.004$ \\
0.5 & $A\to B$ & $1.963\pm0.027$ & $0.899\pm0.013$ & $0.891\pm0.012$ \\
0.5 & $B\to A$ & $1.906\pm0.029$ & $0.856\pm0.013$ & $0.849\pm0.013$ \\
1 & $A\to B$ & $8.046\pm0.108$ & $3.687\pm0.049$ & $3.555\pm0.046$ \\
1 & $B\to A$ & $7.843\pm0.116$ & $3.524\pm0.052$ & $3.404\pm0.048$ \\
\bottomrule
\end{tabular}

\caption{Conditional spin information for the three equal-$\dd q$ bins and the $kk$ measurement. In $A\to B$, $A$ generates the events and $B$ is the alternative; the reverse direction interchanges them. The last two columns are percentages. Uncertainties are four empirical standard errors from the paired jackknife; displayed intervals are rounded outward after rounding the centers. They describe integration noise, not experimental errors or calibrated confidence intervals. $f_N$ refers only to the criterion based on the expected log likelihood ratio of Eq.~\eqref{eq:eventfraction}.}
\label{tab:information}\end{table*}

Reducing the spin analyzing power rapidly reduces the additional information. For $\eta=0.5$, $f_N$ is $0.891\%$ or $0.849\%$, and for $\eta=0.25$ it is $0.223\%$ or $0.212\%$. At $\eta=0$, the spin probabilities are identical and $D_S=0$. Replacing the complete spin states by their target-independent leading-power states, while keeping the complete rates, also gives zero to numerical roundoff. These null comparisons isolate the information carried by the correction to the spin state.

The quartic template uses the consistently expanded state $\brho_0+\delta\brho_4$ and the bin rates $T_{hi}=\Tr(\bR_2+\bR_4)$. It gives $D_S$ larger than the full-kernel values by $15.54\%$ and $16.08\%$ at $\eta=1$, with $f_N=4.07\%$ and $3.91\%$. This difference is a truncation effect in both the rate and the spin state, separate from the integration error, and motivates using the complete kernel for the information estimates. The single-bin spin divergence at $q/m=0.2$ agrees within four combined empirical standard errors with the estimate from the refined, statistically independent scrambles. That check does not validate the three-bin shape independently.

\begin{figure}[t]\centering
\includegraphics[width=\columnwidth]{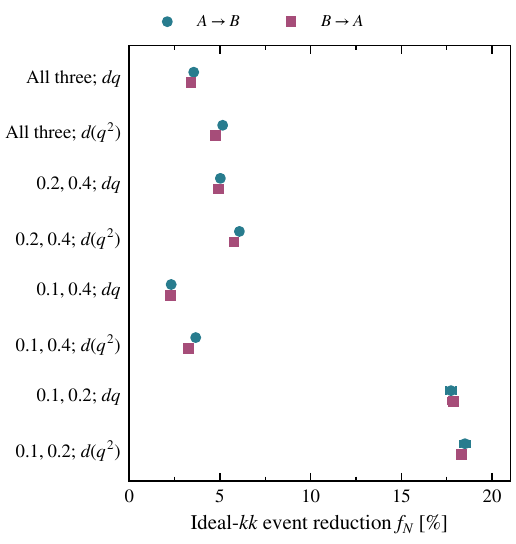}
\caption{Transverse-photon dependence on the retained bins and their measure for an ideal $kk$ measurement. Labels list $q/m$; ``All three'' denotes $0.1,0.2,0.4$. Both directions, all three omissions and both measures are retained. Bars are four paired numerical standard errors. The horizontal coordinate has the interpretation in terms of the expected log likelihood ratio of Eq.~\eqref{eq:eventfraction}, rather than fixed-power or luminosity savings.}
\label{fig:cellchoice}\end{figure}

All three leave-one-bin-out choices and both bin measures were specified before evaluating the information quantities. Their ideal-$kk$ values of $f_N$ span approximately $2.3\%$--$18.5\%$ (Fig.~\ref{fig:cellchoice}). The larger relative gains occur when the retained kinematic shape is less discriminating; they do not imply a larger absolute spin signal or an optimized experimental binning. Equal allocation to the three diagonal spin settings gives $f_N=1.80\%,1.73\%$ for the original three bins. Equal allocation to all nine Pauli setting pairs gives $0.647\%,0.620\%$. These are different fixed measurement allocations, not a proof that $kk$ is optimal or that tomography is intrinsically less informative.

The known-normalization unpolarized information in Eq.~\eqref{eq:knownnorm} is $0.342686$ or $0.222362$, much larger than the shape-only information. Profiling therefore deliberately removes most of the rate discrimination. Even after that removal, the ideal spin increment is modest for the three-bin comparison. Assessing the gain after refitting target shape parameters, screening, fragmentation or detector response requires a joint analysis.

The uncertainty calculation deletes each of the eight common scramble seeds simultaneously across both targets and all three bins. It includes covariance between rates and states. No central matrix is projected or clipped. All positive-$\eta$ increments in the full-kernel comparisons exceed four empirical standard errors; the denominators of quoted ratios are resolved by the same criterion. Analytic synthetic states, deliberate input corruptions, and a separate high-precision calculation from diagonal populations check the statistical assembly. Their scope and shared physical inputs are documented in Appendix~\ref{app:informationchecks}.

\subsection{Limitations of the model comparison}
A target fit would have to retain the unpolarized constraints while allowing the relevant shape parameters to vary. Dipole descriptions of inclusive DIS range from the saturation and impact-parameter models of Refs.~\cite{Golec-Biernat:1998zce,Kowalski:2003hm} to evolved fits including heavy flavors \cite{Albacete:2010sy}. The combined HERA measurements \cite{H1:2015ubc}, Bayesian initial-condition inference \cite{Casuga:2023dcf}, and synthetic inverse-reconstruction studies \cite{Hanninen:2025iuv} illustrate distinct data and model restrictions. A recent nuclear dipole extraction combines a neural-network representation with collinearly improved BK evolution and constraints from forward-hadron production and coherent $J/\psi$ photoproduction \cite{Dai:2026rgl}. The two Gaussian templates used here are not fits to those data. Allowing additional nuisance directions could reduce the residual spin discrimination, and requires a separate joint analysis.

The fixed Gaussian templates also omit evolution. The Balitsky and JIMWLK hierarchies evolve Wilson-line operators \cite{Balitsky:1995ub,Jalilian-Marian:1997qno,Iancu:2000hn}, with the BK equation providing a closed dipole approximation \cite{Kovchegov:1999yj}. Even within dipole evolution, running-coupling prescriptions \cite{Balitsky:2006wa,Kovchegov:2006vj} and the full NLO kernel \cite{Balitsky:2007feb} modify the dynamics. Resummation addresses the large double transverse logarithms \cite{Iancu:2015vea}, and formulations in target rapidity organize these corrections for dilute--dense scattering \cite{Ducloue:2019ezk}. Comparisons of improved BK schemes show that different prescriptions can give comparable descriptions of inclusive HERA data \cite{Xiang:2022iuh}.

Selected quadrupole geometries can be well described by Gaussian closure \cite{Dumitru:2011vk}, but this agreement is not an operator identity. Recent finite-$N_c$ initial-condition studies \cite{Cougoulic:2026drr}, evolved-TMD comparisons \cite{Cougoulic:2026sww}, and explicit JIMWLK counterexamples \cite{Lappi:2026kxi} illustrate the limits of using a fixed Gaussian model to describe general multipoint correlators evolved in QCD. Beyond-Gaussian constructions are available \cite{Penttala:2025tmp}; their spin matrices have not been evaluated here.

Radiative corrections can also change the conditioning information. Treatments that separate Sudakov and small-$x$ logarithms \cite{Mueller:2012uf,Xiao:2017yya} include detailed one-loop derivations of Sudakov double logarithms for heavy-pair and DIS dijet production \cite{Mueller:2013wwa}. In a dilute quark-target calculation, the gluon TMD satisfies both Collins--Soper and BFKL evolution at leading logarithmic accuracy \cite{Zhou:2016tfe}. NLO DIS dijet calculations \cite{Caucal:2021ent,Caucal:2022ulg,Caucal:2023nci,Caucal:2023fsf} and unified evolution analyses \cite{Caucal:2024bae,Caucal:2025mth} extend the treatment of radiation. Ref.~\cite{Caucal:2025mth} extends the treatment of real-emission contributions at leading power.

Sudakov resummation changes back-to-back momentum correlations, as illustrated by a Gaussian target calculation with running-coupling BK evolution for spin-summed DIS dijets \cite{Zhao:2021kae}. Final-state soft radiation itself generates a $\cos 2\phi$ harmonic \cite{Hatta:2020bgy}; its separation from the linearly polarized gluon contribution has been studied in a resummed DIS calculation \cite{Hatta:2021jcd}. A recent proposal defines the recoil from hadrons in a fiducial rapidity interval to reduce both broadening and competing soft harmonics \cite{Lin:2026typ}. Such changes to the momentum distribution would also change the unpolarized information used to condition a spin comparison.

Dihadron observables introduce fragmentation-dependent logarithms \cite{Caucal:2024nsb}; heavy-meson TMD factorization and phenomenology have also been developed in a boosted regime \cite{delCastillo:2020omr,delCastillo:2021znl}. For heavy-flavor dijet Sivers asymmetries, mass-dependent jet and collinear-soft functions have been computed at next-to-leading order and used in next-to-leading-logarithmic resummation \cite{Kang:2020xgk}. Soft-gluon resummation for heavy-meson pair photoproduction has also been formulated to interpolate between massive and massless limits \cite{Marquet:2025jdr}. Extending Eq.~\eqref{eq:kl} requires a treatment of radiation that retains the finite-mass joint final-state spin matrix. The longitudinal cancellation derived here applies at tree level; its behavior after unresolved radiation requires a separate calculation.

Measuring the quark spin requires an analyzer in its hadronic final state. Heavy-baryon polarization retention depends on production and decay \cite{Falk:1993rf,Galanti:2015pqa}; realistic heavy-quark spin-correlation proposals require rate and reconstruction studies \cite{Kats:2023zxb}. Polarized heavy-hadron TMD fragmentation \cite{vonKuk:2023jfd} and the established single- and dihadron fragmentation channels \cite{Collins:1992kk,Boer:1997mf,Boer:2003ya,Metz:2016swz} provide possible inputs for a spin-response model. The value of $\eta$ for this observable still requires calibration.

Correlated hadron spins have long been proposed as probes of polarized fragmentation. Chen \emph{et al.} formulated the production, fragmentation and weak-decay density matrices for spin-$1/2$ baryon pairs in unpolarized $e^+e^-$ annihilation \cite{Chen:1994ar}. More recent work develops longitudinal dihadron correlations in unpolarized $e^+e^-$ and $pp$ collisions \cite{Zhang:2023ugf}, and transverse correlations, including photon-induced production, as probes of spin-transfer fragmentation functions \cite{Yang:2024kjn}. With a thrust selection, polarized fragmenting jet functions allow next-to-next-to-leading-logarithmic predictions for $\Lambda\bar\Lambda$ spin correlations in $e^+e^-$ annihilation \cite{Lin:2025eci}. Studies of nearby hadrons from a single parton and the light-spin state in excited heavy baryons also apply quantum-information constraints to hadronization \cite{vonKuk:2025kbv}. Applying these methods to the separated heavy-quark pair considered here requires a description of its hadronic spin response in addition to the perturbative pair density matrix.

The production region and decay channel affect the measured spin correlation. In a model of fixed-target semi-inclusive DIS, target fragmentation can dilute longitudinal spin transfer to $\Lambda$ hyperons \cite{Zhao:2024usu}. Helicity correlations between hadrons in the current and target regions involve both polarized fragmentation and fracture functions \cite{Xi:2025feb}. Weak hyperon decays then convert the hadronic spin state into an angular distribution, as described in the production-and-decay review of Ref.~\cite{Schonning:2023mge} and the generalized quantum-measurement formulation of Ref.~\cite{Wu:2024mtj}. These results motivate specifying the production region and decay analyzer. The corresponding heavy-flavor response factor requires a channel-specific determination.

Dihadron Bell proposals \cite{Cheng:2025cuv} and recent heavy-quark tomography \cite{FucillaTomography2026} address the connection between spin states and measurements more directly. The latter treats finite-mass kinematics with leading-twist fragmentation functions while leaving genuine higher-power fragmentation effects unresolved. Some of its spin-dependent fragmentation inputs still require quantitative determination. An experimental application of the present calculation would therefore need a specified channel, its spin response, event-axis reconstruction, acceptance and available yields.

\section{Conclusions}
Within the homogeneous Gaussian model, the complete quartic correction restores target dependence of the azimuthally averaged transverse heavy-quark spin state in finite-mass small-$x$ DIS. The calculation retains explicit spin indices, finite $N_c$, all quartic interference terms, self-dipole logarithms and the contact distribution. In the established amplitude-level separation of twists, additional current insertions account for most of the target contrast at the sampled points. The one-current kinematic contribution is smaller but not zero.

The full-kernel calculations cover 28 distinct parameter points and satisfy the stated sampling and truncation criteria. The infrared and kinematic study resolves state changes in all 12 target pairs and negativity changes in 11. Both states in the larger-mass pair remain separable, despite a resolved spin-correlation difference. Changing the screening scale by a factor of two in either direction changes the central state-distance amplitude by $+2.11\%$ or $-3.42\%$ at fixed implicit $Q_s$. The refined $P/m_0=4$ calculations have four-standard-error precision fractions below $3.6\%$ of the quartic effect. These comparisons test discrete points within the Gaussian prescription.

The longitudinal channel and the large-$P$ transverse limit provide analytic checks. Consistent with the earlier longitudinal maximal-entanglement results \cite{Qi2026,FucillaDiffractive2026,Fucilla2026}, for longitudinal photons the spin vector factorizes for arbitrary spin-independent eikonal scattering at fixed kinematics; the prescribed angular average preserves its maximal entanglement. In the transverse channel, normalization cancels the leading quartic vector contribution at large $P$, leaving mass interference generically proportional to $m\ln P/P^3$ at fixed $Q^2$. The latter is an asymptote of the calculated quartic term, not a full-kernel result at $P/m_0>4$.

The conditional-likelihood calculation distinguishes target dependence from added information. After profiling one overall normalization, the three-bin unpolarized shapes still discriminate between the two Gaussian targets. For a purely transverse photon, an ideal $kk$ spin measurement adds $3.5$--$3.7\%$ to their expected log-likelihood information, corresponding to $3.4$--$3.6\%$ fewer selected events at the same expectation. A correlation analyzing factor of $0.25$ reduces the latter fraction to about $0.22\%$. The gain also depends on the retained bins, and the quartic approximation overestimates the primary spin increment by about $16\%$. The increment is small and measurement dependent for these fixed templates.

A continuous joint target fit and a calibrated hadronic spin response are needed to determine whether that increment survives more realistic unpolarized constraints and nuisance parameters. Evolution, radiative corrections and finite-bin acceptance can alter both parts of the likelihood. Independent multipoint sensitivity additionally requires extending the target beyond Gaussian closure. Such extensions require the spin density matrix: a relative correction to the unpolarized cross section does not determine the correction to the normalized spin state.

\appendix
\section{Spin phases and Fourier derivatives}\label{app:spin}
In Eq.~\eqref{eq:axes}, let $\cos\theta=(2z-1)/\beta$ with
$\beta=\sqrt{1-4m^2/M^2}$. Two-component spinors are the columns of
$\chi=R_y(\theta)e^{-i\pi\sigma_z/4}$ and
$\eta=-i\sigma_y\chi^*$ for the antiparticle. Standard Dirac spinors constructed from them, followed by $S(b)$ in Eq.~\eqref{eq:boost}, define all phases. In this representation
$\bar u\{1,\gamma_5,\gamma^i\}\gamma^+v$ is invariant under the common transverse light-front boost. This is why the coefficients of the affine numerator in Eq.~\eqref{eq:wavefunction} are independent of the imbalance direction in the transported basis. The convention includes momentum-dependent spin transport and does not require Wigner rotations to vanish in general.

At fixed external spinors, put $F(k)=(k^2+\eps^2)^{-1}$ and $D=k^2+\eps^2$. The ordinary derivatives are
\begin{align}
 F_i&=-2k_i/D^2,\qquad
 F_{ij}=-2\delta_{ij}/D^2+8k_ik_j/D^3,\\
 F_{ijk}&=\frac{8(\delta_{ij}k_k+\delta_{ik}k_j+\delta_{jk}k_i)}{D^3}
 -\frac{48k_ik_jk_k}{D^4}.
\end{align}
For an affine numerator $N(k)$, $h_I=NF_I+\sum_{a\in I}N_a F_{I\setminus a}$, including repeated-index multiplicities. Under Fourier transformation, a term proportional to $u_I w_J$ contributes
$i^{|I|}(-i)^{|J|}\avg{h_I h_J^\dagger}$ to the spin matrix. Here $I$ and $J$ are lists of transverse indices, $|I|$ and $|J|$ count their entries, and $h_I$ denotes the corresponding derivative of the photon wave function. For the smooth regulated operator expansion, with contracted indices understood, this yields
\begin{align}
 R_3&=-\frac{i}{2}\avg{h_1h_2^\dagger}\mathcal G_{12}
       +\frac{i}{2}\avg{h_2h_1^\dagger}\mathcal G_{21},\\
 R_4&=-\frac16\avg{h_1h_3^\dagger}\mathcal G_{13}
       -\frac16\avg{h_3h_1^\dagger}\mathcal G_{31}\nonumber\\
 &\quad+\frac14\avg{h_2h_2^\dagger}\mathcal G_{22},\\
 \mathcal G_{nm}(q)&=\int\dd^2b\,e^{-iqb}
 \frac{\left\langle\Tr[B_n(v)B_m^\dagger(v')]\right\rangle_A}{\Nc}.
\end{align}
The logarithmic terms in Eq.~\eqref{eq:k4} extend this notation to the nonanalytic MV covariance. They must not be replaced by ill-defined derivatives of a self-dipole at zero separation.

For fixed-angle evaluation, finite Fourier coefficients $F_m(r)$ of the target-direction polynomial are integrated using
\begin{equation}
 \int_0^{2\pi}\dd\vartheta\,e^{-iqr\cos(\vartheta-\phi)}
 e^{im\vartheta}=2\pi(-i)^mJ_m(qr)e^{im\phi}.
\end{equation}
Sixteen uniformly spaced directions exactly resolve the allowed cubic and quartic harmonics; a 24-direction evaluation tests the implementation together with tighter radial integration. Linear distributional terms are supplied by the dilute amplitude rather than sampled at $b=0$.

\section{Complete radial coefficients}\label{app:coeff}
Here $r=|b|$, $x=\Lambda r$, $c=\kappa/(4\pi)$, $D=\Nc d(r)$, and $f,g,h,j$ have their definitions in the main text. Define
\begin{align}
 \alpha&=cK_0(x),\qquad \beta=-cxK_1(x),\nonumber\\
 \mathcal L&=2\alpha+\beta=\nabla^2d,\qquad d'=crK_0(x),\nonumber\\
 \mathcal L'&=c\Lambda[xK_0(x)-2K_1(x)],\nonumber\\
 X&=\alpha(\alpha+\beta),\qquad Y=\beta^2/8 .
\end{align}
Let $\boldsymbol v=(3,-2,-4)$, $\boldsymbol w_0=(1,0,-2)$ and
$c_0=2\ln2-2\gamma_E+1$. The coefficients in Eq.~\eqref{eq:invariants} are
\begin{widetext}
\begin{align}
 \frac{(A,B,C)}{\CF}
 ={}& f(\nabla^4d)_{r>0}
       (c_3/16,-t^2/32,-t^2/16)\nonumber\\
 &-\frac{\Nc h}{2}
 \left[\frac{t^2d'\mathcal L'}{16}\boldsymbol v
       +\frac{s_z}{2}(X\boldsymbol w_0+Y\boldsymbol v)\right]\nonumber\\
 &+\frac{\Nc^2j}{8}t^2(d')^2
       \left[\frac{\alpha}{2}\boldsymbol w_0+\frac{\beta}{8}\boldsymbol v\right]\nonumber\\
 &+\left(\frac g{2\Nc}-\frac h\Nc\right)(0,Y,\mathcal L^2/4)
       +(L/\CF)c_0(1,0,0),\label{eq:radialabc}\\
 L={}&\CF\left(\frac h{2\Nc}-\CF g\right)
          \frac{\mathcal L}{2}\frac{\kappa}{16\pi},\qquad
 (\nabla^4d)_{r>0}=c\Lambda^2[4K_0(x)-xK_1(x)] .
 \label{eq:radial}
\end{align}
\end{widetext}
In the first term only, replace $f$ by $f-1$ to define
$(A,B,C,L)_{\rm nonlinear}$. The Fourier coefficients used in Eq.~\eqref{eq:r4} are
\begin{align}
 (a_4,b_4,c_4,\ell_4)
 ={}&2\pi\int_0^\infty r\,\dd r\,J_0(qr)
       (A,B,C,L)_{\rm nonlinear}\nonumber\\
 &-\frac{\CF\kappa q^4}{(q^2+\Lambda^2)^2}
       (c_3/16,-t^2/32,-t^2/16,0).
 \label{eq:integrated}
\end{align}
The last line incorporates the contact distribution. The $c$ in the definition of $\alpha$ is a scalar normalization and is distinct from $c_4$. At small $D$, series expansions of $f,g,h,j$ avoid cancellation in their closed exponential forms. Equations~\eqref{eq:radialabc}--\eqref{eq:integrated} and the exact hard transforms suffice to reproduce the averaged quartic result without the full-kernel integrator.

\subsection{Expansion of the color matrix}
The coefficients in Eq.~\eqref{eq:k4} can also be obtained from an evolution equation for the two singlet components. Introduce a formal parameter $\xi$ that counts dipole degree, keeping the self-dipole logarithms in $s_2$:
\begin{align*}
 s&=\xi^2s_2,\qquad
 \delta=\xi^2\delta_2+\xi^3\delta_3+\xi^4\delta_4,\\
 W&=2D/\Nc+\xi W_1+\xi^2W_2.
\end{align*}
The parameter $\xi$ grades the expansion and does not rescale the arguments of the self-dipole logarithms. Solve $\partial_\tau X=\mathsf M X$ with $X(0)=(1,0)^T$ and evaluate $(1,\sqrt{N_c^2-1})X(1)$. At zeroth order the second singlet component has damping $e^{-D\tau}$; each insertion then gives one of the moments $f,h,j$ or $g=f-h$. The two off-diagonal insertions contribute $\delta_2^2 C_F g/(2N_c)$, while the diagonal insertion in the second component contributes $-\delta_2^2 C_F h/N_c$. Subtracting the two dipoles leaves the disconnected term $C_F^2d(u)d(w)$. These terms reproduce Eq.~\eqref{eq:k4} without differentiating the nonanalytic self-dipoles at the origin.

A separate matrix-power expansion checks these coefficients exactly for $N_c=2,3,5$ through eighth order in $D$, at a fixed nonzero rational choice of $s_2,\delta_2,\delta_3,\delta_4,W_1,W_2$. Direct coefficient extraction from the matrix exponential at nonzero $D$ agrees when the working precision and extraction radius are refined. These checks take the Gaussian color matrix in Eq.~\eqref{eq:color} as their common input.

\section{Comparison with previous work}\label{app:source}
For the large-$N_c$ comparison with Dumitru and Skokov~\cite{Dumitru2016}, set
$\Gamma=2d$ and $D=\CF\Gamma$ in that limit. The equation numbers below refer to arXiv:1605.02739v3. Its symmetric dipole center agrees with the present center at $z=1/2$. The parameter multiplying its explicit self-dipole logarithm is
$Q_{s,\mathrm{paper}}^2=\CF\kappa/(2\pi)$, which differs from our implicit definition. With this mapping, Eq.~\eqref{eq:k4} reduces term by term to its Eq.~(35); the coefficients of the dipole logarithms match its Eqs.~(60),(61), and the disconnected double-log term matches Eq.~(62). The finite-$N_c$ color construction also approaches the large-$N_c$ kernel with the expected $1/N_c^2$ difference.

A factor-of-two difference remains in the transverse leading-log coefficients of Ref.~\cite{Dumitru2016}. Using the matched coordinate logarithm, that paper's definition of the modified WW tensor and its leading-log replacement
$\ell_u\to2\ln[D_P/(\eps\Lambda)]$, we obtain a hard contraction proportional to
\begin{equation}
 \frac{Q_{s,\mathrm{paper}}^2}{8}
 \ln\frac{D_P}{\eps\Lambda}\,
 \widetilde G_{ij}
 \left[\partial_i\nabla_P^2\Phi\,\partial_j\Phi^*
       +\partial_i\Phi\,\partial_j\nabla_P^2\Phi^*\right].
\end{equation}
The scalar derivative reproduces the longitudinal results in Eqs.~(65) and (66) of that version. The massless vector derivative gives twice the transverse coefficients in Eqs.~(63) and (64). For example, at $P^2=\eps^2$, the isotropic coefficient after factoring out $Q_{s,\mathrm{paper}}^2\ln[D_P/(\eps\Lambda)]$ is $-1/(2\eps^2)$ from the vector derivatives, compared with $-1/(4\eps^2)$ in the reference. The comparison keeps $P/\eps$ fixed and retains the leading infrared logarithm. The origin of this coefficient difference remains to be clarified. Our finite-mass calculation uses the exact transforms in Eqs.~\eqref{eq:scalarlog} and \eqref{eq:logs}, checked by direct coordinate-space integration.

\subsection{Coordinate-space check of the logarithmic term}
The logarithmic coefficients compared above appear in Appendix B of arXiv:1605.02739v3,
added in 2018. The \href{https://link.aps.org/accepted/10.1103/PhysRevD.94.014030}{APS-hosted 2016 accepted manuscript}
contains only the useful-integrals appendix. The comparison here is restricted to the specified preprint version.
We check the logarithmic hard factor by direct coordinate-space integration, implemented without using the hard-factor or quartic-expansion routines. With Fourier measure
$\dd^2u/(2\pi)$ and phase $e^{-iP\cdot u}$, use
\begin{align}
 \psi_L(u)&=K_0(\eps u),\quad
 \psi_{T,a}(u)=i\eps K_1(\eps u)\hat u_a,\nonumber\\
 \Phi_L&=D_P^{-1},\qquad\Phi_{T,a}=P_aD_P^{-1}.
\end{align}
Overall wave-function factors cancel in each relative coefficient. Define
\begin{align}
 A_{i,a}&=\int\frac{\dd^2u}{2\pi}e^{-iP\cdot u}u_i\psi_a(u)
              =i\partial_i\Phi_a,\\
 B_{i,a}&=\int\frac{\dd^2u}{2\pi}e^{-iP\cdot u}u_i u^2\psi_a(u)
              =-i\partial_i\nabla_P^2\Phi_a,\\
 B^{\log}_{i,a}(\Lambda)&=\int\frac{\dd^2u}{2\pi}
 e^{-iP\cdot u}u_i u^2\ln\frac1{u^2\Lambda^2}\psi_a(u).
\end{align}
The isotropic and anisotropic projections are respectively the trace and
the $xx-yy$ difference of the two-dimensional hard tensor, for $P=(P,0)$.
The relative leading-log coefficient, with
$Q_{s,\mathrm{paper}}^2\ln[D_P/(\eps\Lambda)]$ removed, is
\begin{equation}
 c_{\cal P}=-\frac{{\cal P}\!\left[BA^\dagger+AB^\dagger\right]}
 {8{\cal P}\!\left[AA^\dagger\right]}.
 \label{eq:independentlog}
\end{equation}
This expression follows from the coefficient
$-Q_{s,\mathrm{paper}}^2/16$ of
$\widetilde G_{ij}u_iw_j[u^2\ell_u+w^2\ell_w]$ after matching the
two WW projections in Eqs.~(67),(68) of the source. Both logarithmic
interference terms in its Eqs.~(60),(61) are included.

Denoting the isotropic and anisotropic projections by $0$ and $2$, ordinary differentiation gives
\begin{align}
 c_T^{(0)}&=-\frac{2\eps^2[(P^2-\eps^2)^2+2P^4]}
 {(P^2+\eps^2)^2(P^4+\eps^4)},\\
 c_T^{(2)}&=c_L^{(0)}=c_L^{(2)}
 =-\frac{2(2\eps^2-P^2)}{(P^2+\eps^2)^2}.
\end{align}

Direct radial and angular integration is compared with ordinary derivatives
at $P/\eps=0.5,1,2,3$. We refine the angular grid from 256 to 384 nodes, extend the radial endpoint from $40/\eps$ to $55/\eps$, and tighten the integration tolerances. The maximum absolute residual across the coordinate/derivative,
refinement, and finite-log-slope checks is below $7.16\times10^{-13}$
in units where $\eps=1$.
Both transverse coefficients are twice the values in Eqs.~(63) and (64) of the reference, whereas both longitudinal coefficients agree with Eqs.~(65) and (66). At
$P^2=\eps^2$, the transverse isotropic value is
$-1/(2\eps^2)$, compared with $-1/(4\eps^2)$ in the reference.

We also check how the transverse polarization average and the two interference terms enter these relative coefficients.
The transverse polarization average multiplies both tensors in
Eq.~\eqref{eq:independentlog} by $1/2$, leaving their ratio invariant.
Keeping only one of the two interference terms halves both longitudinal and transverse
coefficients, which would also change the longitudinal coefficients that agree with the reference. Finally,
\begin{equation}
 \frac{B^{\log}(\Lambda_1)-B^{\log}(\Lambda_2)}
 {\ln(\Lambda_2/\Lambda_1)}=2B
\end{equation}
is checked with $\Lambda_1/\eps=0.02$, $\Lambda_2/\eps=0.04$ using the
unapproximated logarithmic integrals. This comparison checks the coefficient of the leading infrared logarithm independently of the finite constant term. Our finite-mass calculation uses the exact transforms in Eqs.~\eqref{eq:scalarlog} and \eqref{eq:logs}.

Our Dirac calculation agrees with the reflection-even leading-power spin coefficients and the fully averaged state of Fucilla, Hatta and Xiao~\cite{Fucilla2026}. The comparison uses arXiv:2604.11697v1 and the published article: their tensor decompositions are Eqs.~(14),(76) and Eqs.~(14),(B7), respectively; the photon wave functions are Eqs.~(62)--(66) and Eqs.~(A10)--(A14). Our implementation retains the complete $G_{ij}$ contraction in Eq.~\eqref{eq:lp}, including its $\sin2\phi$ term with zero spin trace, which vanishes in the full angular average. We use the two photon channels separately, with their weights fixed directly by Eqs.~\eqref{eq:wavefunctionL},\eqref{eq:wavefunction}.

\section{Five-dimensional reference integral}\label{app:integrator}
For $q>0$ the $b$-independent terms contribute only to the excluded contact distribution. Write $K_c=Q-S_F(u)S_F(w)$, $b=(r,0)$, $u=U(\cos\alpha,\sin\alpha)$ and $w=W(\cos\beta,\sin\beta)$. Averaging the imbalance direction gives $J_0(qr)$. The remaining common rotation relative to $P=(P,0)$ can be integrated analytically. With the Fourier measure of Eq.~\eqref{eq:logs}, the radial measure and relative-angle measure are
\begin{equation}
 2\pi r\,\dd r\,U\,\dd U\,W\,\dd W\,
 \frac{\dd\alpha}{2\pi}\frac{\dd\beta}{2\pi}.
 \label{eq:refmeasure}
\end{equation}
The six spin matrices are $M_{cc}=\avg{cc^\dagger}$,
$M_{xc}=\avg{b_xc^\dagger+cb_x^\dagger}$,
$M_{yc}=\avg{b_yc^\dagger+cb_y^\dagger}$,
$M_{xx}=\avg{b_xb_x^\dagger}$,
$M_{yy}=\avg{b_yb_y^\dagger}$ and
$M_{xy}=\avg{b_xb_y^\dagger+b_yb_x^\dagger}$, using the affine numerator below Eq.~\eqref{eq:logs}.

Put $s=|u-w|$, $e=(u-w)/s$, $a_u=\hat u\cdot e$, $a_w=\hat w\cdot e$, $n_u=e_x\hat u_y-e_y\hat u_x$, and similarly $n_w$. Define $k_\nu(U)=K_\nu(\eps U)$, $J_n=J_n(Ps)$, $X=(J_0-J_2)/2$ and $Y=(J_0+J_2)/2$. After the $u\leftrightarrow w$ antithetic average, the weights multiplying these six matrices are
\begin{align}
 h_{cc}&=k_0(U)k_0(W)J_0,\nonumber\\
 h_{xc}&=\tfrac{\eps J_1}{2}
 [k_1(U)k_0(W)a_u-k_0(U)k_1(W)a_w],\nonumber\\
 h_{yc}&=\tfrac{\eps J_1}{2}
 [k_1(U)k_0(W)n_u-k_0(U)k_1(W)n_w],\nonumber\\
 h_{xx}&=\eps^2 k_1(U)k_1(W)(a_ua_wX+n_un_wY),\nonumber\\
 h_{yy}&=\eps^2 k_1(U)k_1(W)(a_ua_wY+n_un_wX),\nonumber\\
 h_{xy}&=\tfrac{\eps^2}{2} k_1(U)k_1(W)
 (a_un_w+n_ua_w)(X-Y).
 \label{eq:refweights}
\end{align}
The full integrand is $J_0(qr)K_c\sum_Ah_AM_A$ with measure~\eqref{eq:refmeasure}. Limits at $s=0$ are understood continuously. The real Gaussian kernel allows the additional $b\leftrightarrow-b$ antithetic average used in the code.

Each Sobol point supplies three radial quantiles and two uniform angles. The $U,W$ densities are $p_U(U)=\eps^2 U e^{-\eps U}$. For $r$ we use
\begin{align}
 p_r(r)&=0.85\,a^2r e^{-ar}+0.15\,t^2r e^{-tr},\nonumber\\
 a&=\max(Q_s,q,2\Lambda),\qquad t=\Lambda/2.
\end{align}
The first coordinate selects the mixture component and its shape-two gamma quantile; the next two supply the gamma quantiles for $U,W$. In the numerical weights $K_\nu(\eps U)$ is replaced by $e^{\eps U}K_\nu(\eps U)$ and the common importance factor is $2\pi r/[p_r(r)\eps^4]$. This accounts for all radial Jacobians. Coordinates are placed at 52-bit cell midpoints; the conditional mixture quantile is additionally clipped to $[\epsilon_{\rm mach},1-\epsilon_{\rm mach}]$.

Let $K_{\rm dil}$ be the term linear in the target covariance and $K_n^{\rm nl}=K_n-K_n^{\rm dil}$. The sampled remainder is $K_c-K_{\rm dil}-K_2^{\rm nl}$, or additionally minus $K_4^{\rm nl}$ for the quartic control. Their restored matrices are, respectively,
\begin{align}
 R_{\rm base}^{\rm LP}&=R_{\rm dil}+\bR_2-\bR_2^{\rm dil},\nonumber\\
 R_{\rm base}^{(4)}&=R_{\rm base}^{\rm LP}+\bR_4-\bR_4^{\rm dil}.
\end{align}
Here $R_{\rm dil}$ uses the unexpanded coherent dilute amplitude and the other matrices use the deterministic expansion. Restoration precedes normalization. Both controls share the spin-dependent hard coefficients and the same unexpanded Gaussian kernel; their agreement tests integration consistency rather than independently deriving the physical model.

\section{Checks of the conditional-information analysis}\label{app:informationchecks}
The information analysis uses the spin matrices obtained by integrating the unexpanded Gaussian kernel. The bin measures, omissions, measurement settings, analyzing factors and both hypothesis directions were fixed before the information quantities were evaluated. The hypotheses and the $kk$ setting were chosen with prior knowledge of the original matrices and their target contrasts.

The statistical calculation is checked against analytic diagonal and product states. These include equal spin states with different bin shapes, equal shapes with different spin states, identical hypotheses, and nonzero local polarizations and off-diagonal spin correlations. Independently written probabilities are evaluated at 50 and 80 decimal digits and agree to 40 digits. Floating-point comparisons use absolute tolerance $2\times10^{-12}$ plus relative tolerance $10^{-9}$. The profiled Poisson likelihood is evaluated at its analytic normalization maximum and compared with the conditional likelihood, both for expected counts and for arbitrary integer counts. A further control checks that positive but numerically unresolved shape denominators do not produce quoted ratios.

Seven corrupted inputs test rejection of a nonpositive state, zero trace, a nonfinite entry, missing or duplicate bins, a wrong bin momentum and inconsistent seed order. The numerical comparator also rejects prescribed $10^{-4}$ perturbations. These controls test the statistical calculation with analytically constructed inputs.

For the $kk$ protocol, a separate implementation starts directly from the four diagonal populations of each raw spin matrix and applies a classical independent bit-flip channel with retention probability $(1+\sqrt\eta)/2$. It does not import the routines used to extract Pauli coefficients, read the matrices or evaluate the likelihood. High-precision recomputation of all 80 $kk$ comparisons, including the paired delete-one-scramble errors, agrees with the floating-point analysis to absolute differences below $3.3\times10^{-14}$ for reported metrics and $2.0\times10^{-14}$ for their standard errors. This checks the arithmetic by a different route while sharing the same physical input matrices and measurement definition.

For any reported quantity $g$, deleting the same scramble from every bin and target gives estimates $g_{(-r)}$. We use
\begin{align}
 \sigma_g^2&=\frac{n-1}{n}\sum_{r=1}^n
       \left(g_{(-r)}-\bar g_{(-)}\right)^2,\nonumber\\
 b_g&=(n-1)\left(\bar g_{(-)}-g\right).
\end{align}
Here $n=8$ for the primary three-bin inputs and $n=16$ for the refined central pair. The central value is not bias corrected. In the positive-$\eta$ full-kernel comparisons, the largest $|b_{D_S}|/D_S$ is below $3.9\times10^{-5}$. The same paired jackknife propagates rate--state covariance into the information ratios. These errors measure numerical sampling uncertainty; they are not finite-event measurement errors or a theory-error model. The deterministic quartic template is repeated across scramble slots only to use the same interface. Its zero jackknife variation does not include quadrature or truncation uncertainty.

The comparison using the leading-power spin states keeps the complete unpolarized rates and gives $|D_S|<1.2\times10^{-16}$. The separate central-bin calculation uses disjoint scramble seeds but shares the physical kernel and spin amplitudes. The leading-power and central-bin comparisons test the statistical analysis and numerical integration using the same physical kernel.

\section{Numerical uncertainties}\label{app:uncertainties}

\begin{table}[htbp]\centering\small
\resizebox{\columnwidth}{!}{\begin{tabular}{lrrrrrl}
\toprule
Case & $Q_s$ & $4\sigma/E$ & Trunc. & Ctrl. & $\lambda_{\min}$ & Passed \\
\midrule
Base & 0.25 & 0.94 & 5.17 & 2.01 & 0.0411 & yes \\
Base & 0.50 & 1.04 & 12.06 & 1.82 & 0.0418 & yes \\
IR low & 0.25 & 0.79 & 4.56 & 2.63 & 0.0411 & yes \\
IR low & 0.50 & 1.43 & 12.20 & 2.74 & 0.0418 & yes \\
IR high & 0.25 & 1.14 & 6.27 & 1.58 & 0.0412 & yes \\
IR high & 0.50 & 0.78 & 12.84 & 1.25 & 0.0419 & yes \\
$z$ low & 0.25 & 1.17 & 5.47 & 2.20 & 0.0927 & yes \\
$z$ low & 0.50 & 1.27 & 12.21 & 2.16 & 0.0965 & yes \\
$z$ high & 0.25 & 0.85 & 5.06 & 1.92 & 0.0050 & yes \\
$z$ high & 0.50 & 0.93 & 12.15 & 1.66 & 0.0050 & yes \\
$Q^2$ low & 0.25 & 2.19 & 5.75 & 4.55 & 0.0336 & yes \\
$Q^2$ low & 0.50 & 2.64 & 13.19 & 3.98 & 0.0341 & yes \\
$Q^2$ high & 0.25 & 0.27 & 3.97 & 0.67 & 0.0512 & yes \\
$Q^2$ high & 0.50 & 0.30 & 10.45 & 0.59 & 0.0518 & yes \\
$m$ low & 0.25 & 1.54 & 5.49 & 3.13 & 0.0437 & yes \\
$m$ low & 0.50 & 1.97 & 13.43 & 3.13 & 0.0452 & yes \\
$m$ high & 0.25 & 0.46 & 4.36 & 1.07 & 0.0382 & yes \\
$m$ high & 0.50 & 0.47 & 11.33 & 0.93 & 0.0386 & yes \\
Combined 1 & 0.25 & 0.39 & 3.54 & 1.28 & 0.1173 & yes \\
Combined 1 & 0.50 & 0.74 & 10.13 & 1.45 & 0.1213 & yes \\
Combined 2 & 0.25 & 2.26 & 6.80 & 3.38 & 0.0041 & yes \\
Combined 2 & 0.50 & 1.80 & 13.08 & 2.46 & 0.0041 & yes \\
$P=4$ & 0.25 & 2.99 & 4.57 & 7.94 & 0.0298 & yes \\
$P=4$ & 0.50 & 3.56 & 8.11 & 6.05 & 0.0305 & yes \\
\bottomrule
\end{tabular}
}
\caption{All additional transverse-photon points; case parameters and fixed $m_0$ units are specified in Table~\ref{tab:robustness}. Precision, truncation and control resolution
are percentages of the quartic effect $E$. The last column also includes
nested-sample, control-consistency, positivity and refinement requirements,
and independent legacy agreement where available. All 24 calculations satisfy the stated numerical requirements.}
\end{table}

\section*{Data availability}
The numerical data supporting the findings of this study are available from the corresponding author upon reasonable request.
\raggedright
\bibliographystyle{cpc-local}
\bibliography{references}

@article{Fucilla2026,
  author = {M. Fucilla and Y. Hatta and B.-W. Xiao},
  title = {{Quantum entanglement in electron-nucleus collisions: Role of the linearly polarized gluon distribution}},
  year = {2026},
  journal = {Phys. Rev. D},
  volume = {114},
  pages = {034031},
  doi = {10.1103/6c93-2tgg},
  eprint = {2604.11697},
  archivePrefix = {arXiv}
}

@article{Dominguez2011,
  author = {F. Dominguez and C. Marquet and B.-W. Xiao and F. Yuan},
  title = {{Universality of unintegrated gluon distributions at small $x$}},
  year = {2011},
  journal = {Phys. Rev. D},
  volume = {83},
  pages = {105005},
  doi = {10.1103/PhysRevD.83.105005},
  eprint = {1101.0715},
  archivePrefix = {arXiv}
}

@article{Dumitru2016,
  author = {A. Dumitru and V. Skokov},
  title = {{$\cos(4\phi)$ azimuthal anisotropy in small-$x$ DIS dijet production beyond the leading power TMD limit}},
  year = {2016},
  journal = {Phys. Rev. D},
  volume = {94},
  pages = {014030},
  doi = {10.1103/PhysRevD.94.014030},
  eprint = {1605.02739},
  archivePrefix = {arXiv},
  note = {The Appendix B comparison uses arXiv version 3 (2018)}
}

@article{Altinoluk2019,
  author = {T. Altinoluk and R. Boussarie and P. Kotko},
  title = {{Interplay of the CGC and TMD frameworks to all orders in kinematic twist}},
  year = {2019},
  journal = {JHEP},
    volume = {2019},
    number = {05},
  pages = {156},
  doi = {10.1007/JHEP05(2019)156},
  eprint = {1901.01175},
  archivePrefix = {arXiv}
}

@article{Altinoluk2021,
  author = {T. Altinoluk and C. Marquet and P. Taels},
  title = {{Low-$x$ improved TMD approach to the lepto- and hadroproduction of a heavy-quark pair}},
  year = {2021},
  journal = {JHEP},
    volume = {2021},
    number = {06},
  pages = {085},
  doi = {10.1007/JHEP06(2021)085},
  eprint = {2103.14495},
  archivePrefix = {arXiv}
}

@article{Boussarie2021,
  author = {R. Boussarie and H. M{\"a}ntysaari and F. Salazar and B. Schenke},
  title = {{The importance of kinematic twists and genuine saturation effects in dijet production at the Electron-Ion Collider}},
  year = {2021},
  journal = {JHEP},
    volume = {2021},
    number = {09},
  pages = {178},
  doi = {10.1007/JHEP09(2021)178},
  eprint = {2106.11301},
  archivePrefix = {arXiv}
}

@article{Iancu2012,
  author = {E. Iancu and D. N. Triantafyllopoulos},
  title = {{JIMWLK evolution in the Gaussian approximation}},
  year = {2012},
  journal = {JHEP},
    volume = {2012},
    number = {04},
  pages = {025},
  doi = {10.1007/JHEP04(2012)025},
  eprint = {1112.1104},
  archivePrefix = {arXiv}
}

@misc{FucillaTomography2026,
  author = {M. Fucilla and Y. Guo and Y. Hatta},
  title = {{Quantum tomography at the Electron-Ion Collider}},
  year = {2026},
  eprint = {2609.32800},
  archivePrefix = {arXiv}
}

@misc{Agrawal2026,
  author = {S. Agrawal and M. Zahoor and R. Abir},
  title = {{Soft-Radiation-Induced Decoherence of Heavy-Quark Spin Entanglement at the Electron-Ion Collider}},
  year = {2026},
  eprint = {2606.29944},
  archivePrefix = {arXiv}
}

@article{Peres1996,
  author = {A. Peres},
  title = {{Separability criterion for density matrices}},
  year = {1996},
  journal = {Phys. Rev. Lett.},
  volume = {77},
  pages = {1413--1415},
  doi = {10.1103/PhysRevLett.77.1413},
  eprint = {quant-ph/9604005},
  archivePrefix = {arXiv}
}

@article{Horodecki1996,
  author = {M. Horodecki and P. Horodecki and R. Horodecki},
  title = {{Separability of mixed states: necessary and sufficient conditions}},
  year = {1996},
  journal = {Phys. Lett. A},
  volume = {223},
  pages = {1--8},
  doi = {10.1016/S0375-9601(96)00706-2},
  eprint = {quant-ph/9605038},
  archivePrefix = {arXiv}
}

@article{Vidal2002,
  author = {G. Vidal and R. F. Werner},
  title = {{Computable measure of entanglement}},
  year = {2002},
  journal = {Phys. Rev. A},
  volume = {65},
  pages = {032314},
  doi = {10.1103/PhysRevA.65.032314},
  eprint = {quant-ph/0102117},
  archivePrefix = {arXiv}
}

@article{Wootters1998,
  author = {W. K. Wootters},
  title = {{Entanglement of formation of an arbitrary state of two qubits}},
  year = {1998},
  journal = {Phys. Rev. Lett.},
  volume = {80},
  pages = {2245--2248},
  doi = {10.1103/PhysRevLett.80.2245},
  eprint = {quant-ph/9709029},
  archivePrefix = {arXiv}
}

@article{Qi2026,
  author = {Wei Qi and Zijing Guo and Bo-Wen Xiao},
  title = {{Studying maximal entanglement and Bell nonlocality at an electron-ion collider}},
  journal = {Phys. Rev. D},
  volume = {113},
  pages = {054048},
  year = {2026},
  doi = {10.1103/6ycn-x3yj},
  eprint = {2506.12889},
  archivePrefix = {arXiv}
}

@article{FucillaDiffractive2026,
  author = {Michael Fucilla and Yoshitaka Hatta},
  title = {{Spin-spin entanglement in diffractive heavy-quark production}},
  journal = {Phys. Rev. D},
  volume = {113},
  pages = {L031504},
  year = {2026},
  doi = {10.1103/gbk8-z3dd},
  eprint = {2509.05267},
  archivePrefix = {arXiv}
}

@misc{Kar2026,
  author = {Tiyasa Kar and Swagato Mukherjee and Vladimir V. Skokov and Shaswat Tiwari and Fei Yao},
  title = {{DIS Dijet Production: An operator basis bridging eikonal and TMD regimes}},
  year = {2026},
  eprint = {2609.12363},
  archivePrefix = {arXiv}
}

@article{KarMethods2026,
  author = {Tiyasa Kar and Andrey Tarasov and Vladimir V. Skokov},
  title = {{DIS dijet production in background field approach: General formalism and methods}},
  year = {2026},
  journal = {Phys. Rev. D},
  volume = {114},
  pages = {034010},
  doi = {10.1103/8ggz-1434},
  eprint = {2603.08805},
  archivePrefix = {arXiv}
}

@article{McLerran:1993ni,
  author = {McLerran, Larry and Venugopalan, Raju},
  title = {{Computing quark and gluon distribution functions for very large nuclei}},
  eprint = {hep-ph/9309289},
  archivePrefix = {arXiv},
  primaryClass = {hep-ph},
  doi = {10.1103/PhysRevD.49.2233},
  journal = {Phys. Rev. D},
  volume = {49},
  pages = {2233--2241},
  year = {1994}
}

@article{McLerran:1993ka,
  author = {McLerran, Larry and Venugopalan, Raju},
  title = {{Gluon distribution functions for very large nuclei at small transverse momentum}},
  eprint = {hep-ph/9311205},
  archivePrefix = {arXiv},
  primaryClass = {hep-ph},
  doi = {10.1103/PhysRevD.49.3352},
  journal = {Phys. Rev. D},
  volume = {49},
  pages = {3352--3355},
  year = {1994}
}

@article{Balitsky:1995ub,
  author = {Balitsky, I.},
  title = {{Operator expansion for high-energy scattering}},
  eprint = {hep-ph/9509348},
  archivePrefix = {arXiv},
  primaryClass = {hep-ph},
  doi = {10.1016/0550-3213(95)00638-9},
  journal = {Nucl. Phys. B},
  volume = {463},
  pages = {99--157},
  year = {1996}
}

@article{Kovchegov:1999yj,
  author = {Kovchegov, Yuri V.},
  title = {{Small-$x$ $F_{2}$ structure function of a nucleus including multiple Pomeron exchanges}},
  eprint = {hep-ph/9901281},
  archivePrefix = {arXiv},
  primaryClass = {hep-ph},
  doi = {10.1103/PhysRevD.60.034008},
  journal = {Phys. Rev. D},
  volume = {60},
  pages = {034008},
  year = {1999}
}

@article{Jalilian-Marian:1997qno,
  author = {Jalilian-Marian, Jamal and Kovner, Alex and Leonidov, Andrei and Weigert, Heribert},
  title = {{The BFKL equation from the Wilson renormalization group}},
  eprint = {hep-ph/9701284},
  archivePrefix = {arXiv},
  primaryClass = {hep-ph},
  doi = {10.1016/S0550-3213(97)00440-9},
  journal = {Nucl. Phys. B},
  volume = {504},
  pages = {415--431},
  year = {1997}
}

@article{Iancu:2000hn,
  author = {Iancu, Edmond and Leonidov, Andrei and McLerran, Larry},
  title = {{Nonlinear gluon evolution in the color glass condensate: I}},
  eprint = {hep-ph/0011241},
  archivePrefix = {arXiv},
  primaryClass = {hep-ph},
  doi = {10.1016/S0375-9474(01)00642-X},
  journal = {Nucl. Phys. A},
  volume = {692},
  pages = {583--645},
  year = {2001}
}

@article{Gelis:2010nm,
  author = {Gelis, Francois and Iancu, Edmond and Jalilian-Marian, Jamal and Venugopalan, Raju},
  title = {{The Color Glass Condensate}},
  eprint = {1002.0333},
  archivePrefix = {arXiv},
  primaryClass = {hep-ph},
  doi = {10.1146/annurev.nucl.010909.083629},
  journal = {Ann. Rev. Nucl. Part. Sci.},
  volume = {60},
  pages = {463--489},
  year = {2010}
}

@article{Fukushima:2007dy,
  author = {Fukushima, Kenji and Hidaka, Yoshimasa},
  title = {{Light projectile scattering off the Color Glass Condensate}},
  eprint = {0704.2806},
  archivePrefix = {arXiv},
  primaryClass = {hep-ph},
  doi = {10.1088/1126-6708/2007/06/040},
  journal = {JHEP},
  volume = {2007},
  number = {06},
  pages = {040},
  year = {2007}
}

@article{Fukushima:2017mko,
  author = {Fukushima, Kenji and Hidaka, Yoshimasa},
  title = {{General formulae for dipole Wilson line correlators with the Color Glass Condensate}},
  eprint = {1708.03051},
  archivePrefix = {arXiv},
  primaryClass = {hep-ph},
  doi = {10.1007/JHEP11(2017)114},
  journal = {JHEP},
    volume = {2017},
  pages = {114},
  year = {2017},
    number = {11}
}

@article{Dumitru:2011vk,
  author = {Dumitru, A. and Jalilian-Marian, J. and Lappi, T. and Schenke, B. and Venugopalan, R.},
  title = {{Renormalization group evolution of multi-gluon correlators in high energy QCD}},
  eprint = {1108.4764},
  archivePrefix = {arXiv},
  primaryClass = {hep-ph},
  doi = {10.1016/j.physletb.2011.11.002},
  journal = {Phys. Lett. B},
  volume = {706},
  pages = {219--224},
  year = {2011}
}

@article{Penttala:2025tmp,
  author = {Penttala, Jani},
  title = {{Color-glass condensate beyond the Gaussian approximation}},
  eprint = {2507.18711},
  archivePrefix = {arXiv},
  primaryClass = {hep-ph},
  doi = {10.1007/JHEP02(2026)059},
  journal = {JHEP},
    volume = {2026},
  pages = {059},
  year = {2026},
    number = {02}
}

@article{Lappi:2026kxi,
  author = {Lappi, T. and Triantafyllopoulos, D. N.},
  title = {{When JIMWLK evolution really matters: The example of incoherent diffraction}},
  eprint = {2604.26123},
  archivePrefix = {arXiv},
  primaryClass = {hep-ph},
  doi = {10.1103/m4h9-s6y6},
  journal = {Phys. Rev. D},
  volume = {114},
  pages = {014036},
  year = {2026}
}

@article{Agostini:2025vvx,
  author = {Agostini, Pedro and Altinoluk, Tolga and Armesto, N{\'{e}}stor and Beuf, Guillaume and Cougoulic, Florian and Mulani, Swaleha},
  title = {{Dijet production in DIS off a large nucleus at next-to-eikonal accuracy in a Gaussian model within the CGC framework}},
  eprint = {2512.17848},
  archivePrefix = {arXiv},
  primaryClass = {hep-ph},
  doi = {10.1103/fpy6-9s8c},
  journal = {Phys. Rev. D},
  volume = {113},
  pages = {054035},
  year = {2026}
}

@misc{Duan:2026yrk,
  author = {Duan, Haowu and Dai, Si-Wei and Yi, Cong and Zhao, Wenbin},
  title = {{Fourier Transforms of Color Glass Condensate Multi-Wilson-Line Correlators via Filon Quadrature}},
  eprint = {2608.18589},
  archivePrefix = {arXiv},
  primaryClass = {hep-ph},
  year = {2026}
}

@misc{Duan:2026jrk,
  author = {Duan, Haowu and Yi, Cong and Dai, Si-Wei and Wei, Shu-Yi and Zhao, Wenbin and Zheng, Liang},
  title = {{CGC-py: A Monte Carlo Event Generator for Gluon Saturation Physics}},
  eprint = {2609.03434},
  archivePrefix = {arXiv},
  primaryClass = {hep-ph},
  year = {2026}
}

@article{Golec-Biernat:1998zce,
  author = {Golec-Biernat, K. and W{\"{u}}sthoff, M.},
  title = {{Saturation effects in deep inelastic scattering at low $Q^{2}$ and its implications on diffraction}},
  eprint = {hep-ph/9807513},
  archivePrefix = {arXiv},
  primaryClass = {hep-ph},
  doi = {10.1103/PhysRevD.59.014017},
  journal = {Phys. Rev. D},
  volume = {59},
  pages = {014017},
  year = {1998}
}

@article{Kowalski:2003hm,
  author = {Kowalski, Henri and Teaney, Derek},
  title = {{Impact parameter dipole saturation model}},
  eprint = {hep-ph/0304189},
  archivePrefix = {arXiv},
  primaryClass = {hep-ph},
  doi = {10.1103/PhysRevD.68.114005},
  journal = {Phys. Rev. D},
  volume = {68},
  pages = {114005},
  year = {2003}
}

@article{Albacete:2010sy,
  author = {Albacete, Javier L. and Armesto, N{\'{e}}stor and Milhano, Jos{\'{e}} Guilherme and Quiroga Arias, Paloma and Salgado, Carlos A.},
  title = {{AAMQS: A non-linear QCD analysis of new HERA data at small-x including heavy quarks}},
  eprint = {1012.4408},
  archivePrefix = {arXiv},
  primaryClass = {hep-ph},
  doi = {10.1140/epjc/s10052-011-1705-3},
  journal = {Eur. Phys. J. C},
  volume = {71},
  pages = {1705},
  year = {2011}
}

@article{H1:2015ubc,
  author = {Abramowicz, H. and others},
  title = {{Combination of measurements of inclusive deep inelastic ${e^{\pm}p}$ scattering cross sections and QCD analysis of HERA data}},
  eprint = {1506.06042},
  archivePrefix = {arXiv},
  primaryClass = {hep-ex},
  doi = {10.1140/epjc/s10052-015-3710-4},
  journal = {Eur. Phys. J. C},
  volume = {75},
  pages = {580},
  collaboration = {H1 and ZEUS},
  year = {2015}
}

@article{Casuga:2023dcf,
  author = {Casuga, Carlisle and Karhunen, Mikko and M{\"{a}}ntysaari, Heikki},
  title = {{Inferring the initial condition for the Balitsky-Kovchegov equation}},
  eprint = {2311.10491},
  archivePrefix = {arXiv},
  primaryClass = {hep-ph},
  doi = {10.1103/PhysRevD.109.054018},
  journal = {Phys. Rev. D},
  volume = {109},
  pages = {054018},
  year = {2024}
}

@article{Hanninen:2025iuv,
  author = {H{\"{a}}nninen, H. and Kykk{\"{a}}nen, A. and Schl{\"{u}}ter, H.},
  title = {{Reconstruction of the dipole amplitude in the dipole picture as a mathematical inverse problem}},
  eprint = {2509.05005},
  archivePrefix = {arXiv},
  primaryClass = {hep-ph},
  doi = {10.1103/7fhf-7fp4},
  journal = {Phys. Rev. D},
  volume = {112},
  pages = {094026},
  year = {2025}
}

@article{Accardi:2012qut,
  author = {Accardi, A. and others},
  title = {{Electron-Ion Collider: The next QCD frontier: Understanding the glue that binds us all}},
  eprint = {1212.1701},
  archivePrefix = {arXiv},
  primaryClass = {nucl-ex},
  doi = {10.1140/epja/i2016-16268-9},
  journal = {Eur. Phys. J. A},
  volume = {52},
  pages = {268},
  year = {2016}
}

@article{Morreale:2021pnn,
  author = {Morreale, Astrid and Salazar, Farid},
  title = {{Mining for Gluon Saturation at Colliders}},
  eprint = {2108.08254},
  archivePrefix = {arXiv},
  primaryClass = {hep-ph},
  doi = {10.3390/universe7080312},
  journal = {Universe},
  volume = {7},
  pages = {312},
  year = {2021}
}

@article{Aschenauer:2017jsk,
  author = {Aschenauer, E C and Fazio, S and Lee, J H and M{\"{a}}ntysaari, H and Page, B S and Schenke, B and Ullrich, T and Venugopalan, R and Zurita, P},
  title = {{The electron--ion collider: assessing the energy dependence of key measurements}},
  eprint = {1708.01527},
  archivePrefix = {arXiv},
  primaryClass = {nucl-ex},
  doi = {10.1088/1361-6633/aaf216},
  journal = {Rept. Prog. Phys.},
  volume = {82},
  pages = {024301},
  year = {2019}
}

@article{Mulders:2000sh,
    author = "Mulders, P. J. and Rodrigues, J.",
    title = {{Transverse momentum dependence in gluon distribution and fragmentation functions}},
    eprint = "hep-ph/0009343",
    archivePrefix = "arXiv",
    reportNumber = "VUTH-00-23",
    doi = "10.1103/PhysRevD.63.094021",
    journal = "Phys. Rev. D",
    volume = "63",
    pages = "094021",
    year = "2001"
}

@article{Boer:2010zf,
    author = "Boer, Dani{\"e}l and Brodsky, Stanley J. and Mulders, Piet J. and Pisano, Cristian",
    title = {{Direct Probes of Linearly Polarized Gluons inside Unpolarized Hadrons}},
    eprint = "1011.4225",
    archivePrefix = "arXiv",
    primaryClass = "hep-ph",
    reportNumber = "SLAC-PUB-14294",
    doi = "10.1103/PhysRevLett.106.132001",
    journal = "Phys. Rev. Lett.",
    volume = "106",
    pages = "132001",
    year = "2011"
}

@article{Pisano:2013cya,
    author = "Pisano, Cristian and Boer, Dani{\"e}l and Brodsky, Stanley J. and Buffing, Maarten G. A. and Mulders, Piet J.",
    title = {{Linear polarization of gluons and photons in unpolarized collider experiments}},
    eprint = "1307.3417",
    archivePrefix = "arXiv",
    primaryClass = "hep-ph",
    reportNumber = "NIKHEF-2013-023, SLAC-PUB-15688",
    doi = "10.1007/JHEP10(2013)024",
    journal = "JHEP",
    volume = {2013},
    number = {10},
    pages = "024",
    year = "2013"
}

@article{Metz:2011wb,
    author = "Metz, Andreas and Zhou, Jian",
    title = {{Distribution of linearly polarized gluons inside a large nucleus}},
    eprint = "1105.1991",
    archivePrefix = "arXiv",
    primaryClass = "hep-ph",
    doi = "10.1103/PhysRevD.84.051503",
    journal = "Phys. Rev. D",
    volume = "84",
    pages = "051503",
    year = "2011"
}

@article{Dominguez:2011br,
    author = "Dominguez, Fabio and Qiu, Jian-Wei and Xiao, Bo-Wen and Yuan, Feng",
    title = {{Linearly polarized gluon distributions in the color dipole model}},
    eprint = "1109.6293",
    archivePrefix = "arXiv",
    primaryClass = "hep-ph",
    doi = "10.1103/PhysRevD.85.045003",
    journal = "Phys. Rev. D",
    volume = "85",
    pages = "045003",
    year = "2012"
}

@article{Dumitru:2015gaa,
    author = "Dumitru, Adrian and Lappi, Tuomas and Skokov, Vladimir",
    title = {{Distribution of Linearly Polarized Gluons and Elliptic Azimuthal Anisotropy in Deep Inelastic Scattering Dijet Production at High Energy}},
    eprint = "1508.04438",
    archivePrefix = "arXiv",
    primaryClass = "hep-ph",
    doi = "10.1103/PhysRevLett.115.252301",
    journal = "Phys. Rev. Lett.",
    volume = "115",
    number = "25",
    pages = "252301",
    year = "2015"
}

@article{Boer:2016fqd,
    author = {Boer, Dani{\"e}l and Mulders, Piet J. and Pisano, Cristian and Zhou, Jian},
    title = {{Asymmetries in heavy quark pair and dijet production at an EIC}},
    eprint = "1605.07934",
    archivePrefix = "arXiv",
    primaryClass = "hep-ph",
    doi = "10.1007/JHEP08(2016)001",
    journal = "JHEP",
    volume = {2016},
    number = {08},
    pages = "001",
    year = "2016"
}

@article{Marquet:2016cgx,
    author = "Marquet, C. and Petreska, E. and Roiesnel, C.",
    title = {{Transverse-momentum-dependent gluon distributions from JIMWLK evolution}},
    eprint = "1608.02577",
    archivePrefix = "arXiv",
    primaryClass = "hep-ph",
    doi = "10.1007/JHEP10(2016)065",
    journal = "JHEP",
    volume = {2016},
    number = {10},
    pages = "065",
    year = "2016"
}

@article{Marquet:2017xwy,
    author = "Marquet, Cyrille and Roiesnel, Claude and Taels, Pieter",
    title = {{Linearly polarized small-x gluons in forward heavy-quark pair production}},
    eprint = "1710.05698",
    archivePrefix = "arXiv",
    primaryClass = "hep-ph",
    doi = "10.1103/PhysRevD.97.014004",
    journal = "Phys. Rev. D",
    volume = "97",
    number = "1",
    pages = "014004",
    year = "2018"
}

@article{Efremov:2017iwh,
    author = "Efremov, A. V. and Ivanov, N. Ya. and Teryaev, O. V.",
    title = {{How to measure the linear polarization of gluons in unpolarized proton using the heavy-quark pair leptoproduction}},
    eprint = "1711.05221",
    archivePrefix = "arXiv",
    primaryClass = "hep-ph",
    doi = "10.1016/j.physletb.2017.12.061",
    journal = "Phys. Lett. B",
    volume = "777",
    pages = "435--441",
    year = "2018"
}

@article{Kotko:2015ura,
    author = "Kotko, P. and Kutak, K. and Marquet, C. and Petreska, E. and Sapeta, S. and van Hameren, A.",
    title = {{Improved TMD factorization for forward dijet production in dilute-dense hadronic collisions}},
    eprint = "1503.03421",
    archivePrefix = "arXiv",
    primaryClass = "hep-ph",
    reportNumber = "CERN-PH-TH-2015-045, CPHT-RR005.0315, IFJPAN-IV-2015-2",
    doi = "10.1007/JHEP09(2015)106",
    journal = "JHEP",
    volume = {2015},
    number = {09},
    pages = "106",
    year = "2015"
}

@article{Altinoluk:2019wyu,
    author = "Altinoluk, Tolga and Boussarie, Renaud",
    title = {{Low x physics as an infinite twist (G)TMD framework: unravelling the origins of saturation}},
    eprint = "1902.07930",
    archivePrefix = "arXiv",
    primaryClass = "hep-ph",
    doi = "10.1007/JHEP10(2019)208",
    journal = "JHEP",
    volume = {2019},
    number = {10},
    pages = "208",
    year = "2019"
}

@article{Boussarie:2020vzf,
    author = "Boussarie, Renaud and Mehtar-Tani, Yacine",
    title = {{Gauge invariance of transverse momentum dependent distributions at small x}},
    eprint = "2001.06449",
    archivePrefix = "arXiv",
    primaryClass = "hep-ph",
    doi = "10.1103/PhysRevD.103.094012",
    journal = "Phys. Rev. D",
    volume = "103",
    number = "9",
    pages = "094012",
    year = "2021"
}

@article{Mueller:2012uf,
    author = "Mueller, A. H. and Xiao, Bo-Wen and Yuan, Feng",
    title = {{Sudakov Resummation in the Small-x Saturation Formalism}},
    eprint = "1210.5792",
    archivePrefix = "arXiv",
    primaryClass = "hep-ph",
    doi = "10.1103/PhysRevLett.110.082301",
    journal = "Phys. Rev. Lett.",
    volume = "110",
    number = "8",
    pages = "082301",
    year = "2013"
}

@article{Xiao:2017yya,
    author = "Xiao, Bo-Wen and Yuan, Feng and Zhou, Jian",
    title = {{Transverse momentum dependent parton distributions at small-x}},
    eprint = "1703.06163",
    archivePrefix = "arXiv",
    primaryClass = "hep-ph",
    doi = "10.1016/j.nuclphysb.2017.05.012",
    journal = "Nucl. Phys. B",
    volume = "921",
    pages = "104--126",
    year = "2017"
}

@article{Caucal:2021ent,
    author = "Caucal, Paul and Salazar, Farid and Venugopalan, Raju",
    title = {{Dijet impact factor in DIS at next-to-leading order in the Color Glass Condensate}},
    eprint = "2108.06347",
    archivePrefix = "arXiv",
    primaryClass = "hep-ph",
    doi = "10.1007/JHEP11(2021)222",
    journal = "JHEP",
    volume = {2021},
    number = {11},
    pages = "222",
    year = "2021"
}

@article{Caucal:2022ulg,
    author = {Caucal, Paul and Salazar, Farid and Schenke, Bj{\"o}rn and Venugopalan, Raju},
    title = {{Back-to-back inclusive dijets in DIS at small x: Sudakov suppression and gluon saturation at NLO}},
    eprint = "2208.13872",
    archivePrefix = "arXiv",
    primaryClass = "hep-ph",
    doi = "10.1007/JHEP11(2022)169",
    journal = "JHEP",
    volume = {2022},
    number = {11},
    pages = "169",
    year = "2022"
}

@article{Caucal:2023nci,
    author = {Caucal, Paul and Salazar, Farid and Schenke, Bj{\"o}rn and Stebel, Tomasz and Venugopalan, Raju},
    title = {{Back-to-back inclusive dijets in DIS at small x: gluon Weizsäcker-Williams distribution at NLO}},
    eprint = "2304.03304",
    archivePrefix = "arXiv",
    primaryClass = "hep-ph",
    doi = "10.1007/JHEP08(2023)062",
    journal = "JHEP",
    volume = {2023},
    number = {08},
    pages = "062",
    year = "2023"
}

@article{Caucal:2023fsf,
    author = {Caucal, Paul and Salazar, Farid and Schenke, Bj{\"o}rn and Stebel, Tomasz and Venugopalan, Raju},
    title = {{Back-to-Back Inclusive Dijets in Deep Inelastic Scattering at Small x: Complete NLO Results and Predictions}},
    eprint = "2308.00022",
    archivePrefix = "arXiv",
    primaryClass = "hep-ph",
    doi = "10.1103/PhysRevLett.132.081902",
    journal = "Phys. Rev. Lett.",
    volume = "132",
    number = "8",
    pages = "081902",
    year = "2024"
}

@article{Caucal:2024nsb,
    author = "Caucal, Paul and Salazar, Farid",
    title = {{Dihadron correlations in small-x DIS at NLO: transverse momentum dependent fragmentation}},
    eprint = "2405.19404",
    archivePrefix = "arXiv",
    primaryClass = "hep-ph",
    reportNumber = "INT-PUB-24-022",
    doi = "10.1007/JHEP12(2024)130",
    journal = "JHEP",
    volume = {2024},
    number = {12},
    pages = "130",
    year = "2024"
}

@article{Caucal:2025mth,
    author = "Caucal, Paul and Iancu, Edmond and Salazar, Farid and Yuan, Feng",
    title = {{Gluon splitting at small x: a unified derivation for the JIMWLK, DGLAP and CSS equations}},
    eprint = "2510.08454",
    archivePrefix = "arXiv",
    primaryClass = "hep-ph",
    doi = "10.1007/JHEP03(2026)198",
    journal = "JHEP",
    volume = {2026},
    number = {03},
    pages = "198",
    year = "2026"
}

@article{Mukherjee:2026cte,
    author = "Mukherjee, Swagato and Skokov, Vladimir V. and Tarasov, Andrey and Tiwari, Shaswat and Yao, Fei",
    title = {{Back-to-back dijet production in DIS at arbitrary Bjorken x: TMD gluon distributions to twist-3 accuracy}},
    eprint = "2602.15137",
    archivePrefix = "arXiv",
    primaryClass = "hep-ph",
    doi = "10.1103/nnhn-jvz7",
    journal = "Phys. Rev. D",
    volume = "114",
    number = "3",
    pages = "036022",
    year = "2026"
}

@misc{Caucal:2026yzl,
    author = "Caucal, Paul and Salazar, Farid",
    title = {{Transverse momentum dependent parton distributions at small $x$}},
    eprint = "2610.26004",
    archivePrefix = "arXiv",
    primaryClass = "hep-ph",
    month = "10",
    year = "2026"
}

@article{delCastillo:2020omr,
    author = "del Castillo, Rafael F. and Echevarria, Miguel G. and Makris, Yiannis and Scimemi, Ignazio",
    title = {{TMD factorization for dijet and heavy-meson pair in DIS}},
    eprint = "2008.07531",
    archivePrefix = "arXiv",
    primaryClass = "hep-ph",
    doi = "10.1007/JHEP01(2021)088",
    journal = "JHEP",
    volume = {2021},
    number = {01},
    pages = "088",
    year = "2021"
}

@article{delCastillo:2021znl,
    author = "del Castillo, Rafael F. and Echevarria, Miguel G. and Makris, Yiannis and Scimemi, Ignazio",
    title = {{Transverse momentum dependent distributions in dijet and heavy hadron pair production at EIC}},
    eprint = "2111.03703",
    archivePrefix = "arXiv",
    primaryClass = "hep-ph",
    doi = "10.1007/JHEP03(2022)047",
    journal = "JHEP",
    volume = {2022},
    number = {03},
    pages = "047",
    year = "2022"
}

@article{Caucal:2024bae,
    author = "Caucal, Paul and Iancu, Edmond",
    title = {{Evolution of the transverse-momentum dependent gluon distribution at small x}},
    eprint = "2406.04238",
    archivePrefix = "arXiv",
    primaryClass = "hep-ph",
    doi = "10.1103/PhysRevD.111.074008",
    journal = "Phys. Rev. D",
    volume = "111",
    number = "7",
    pages = "074008",
    year = "2025"
}

@article{Cougoulic:2026drr,
    author = "Cougoulic, Florian and Korcyl, Piotr and Stebel, Tomasz",
    title = {{Small-x TMD distributions initial condition: Nc-dependence and Gaussian approximations}},
    eprint = "2603.15243",
    archivePrefix = "arXiv",
    primaryClass = "hep-ph",
    doi = "10.1007/JHEP08(2026)029",
    journal = "JHEP",
    volume = {2026},
    number = {08},
    pages = "029",
    year = "2026"
}

@misc{Cougoulic:2026sww,
    author = "Cougoulic, Florian and Korcyl, Piotr",
    title = {{Small-x TMD distributions: JIMWLK evolution and the Gaussian truncation}},
    eprint = "2608.30687",
    archivePrefix = "arXiv",
    primaryClass = "hep-ph",
    month = "8",
    year = "2026"
}

@article{Altinoluk:2024zom,
    author = "Altinoluk, Tolga and Beuf, Guillaume and Czajka, Alina and Marquet, Cyrille",
    title = {{Back-to-back dijet production in DIS at next-to-eikonal accuracy and twist-3 gluon TMDs}},
    eprint = "2410.00612",
    archivePrefix = "arXiv",
    primaryClass = "hep-ph",
    doi = "10.1103/PhysRevD.111.014010",
    journal = "Phys. Rev. D",
    volume = "111",
    number = "1",
    pages = "014010",
    year = "2025"
}

@article{Barr:2024djo,
    author = "Barr, Alan J. and Fabbrichesi, Marco and Floreanini, Roberto and Gabrielli, Emidio and Marzola, Luca",
    title = "{Quantum entanglement and Bell inequality violation at colliders}",
    eprint = "2402.07972",
    archivePrefix = "arXiv",
    primaryClass = "hep-ph",
    doi = "10.1016/j.ppnp.2024.104134",
    journal = "Prog. Part. Nucl. Phys.",
    volume = "139",
    pages = "104134",
    year = "2024"
}

@article{Afik:2020onf,
    author = "Afik, Yoav and de Nova, Juan Ram{\'o}n Mu{\~n}oz",
    title = "{Entanglement and quantum tomography with top quarks at the LHC}",
    eprint = "2003.02280",
    archivePrefix = "arXiv",
    primaryClass = "quant-ph",
    doi = "10.1140/epjp/s13360-021-01902-1",
    journal = "Eur. Phys. J. Plus",
    volume = "136",
    number = "9",
    pages = "907",
    year = "2021"
}

@article{Chen:2024qvx,
    author = "Chen, Zhao-Xuan and Dong, Hui and Wei, Shu-Yi",
    title = "{Dihadron helicity correlation in photon-nucleus collisions}",
    eprint = "2404.19202",
    archivePrefix = "arXiv",
    primaryClass = "hep-ph",
    doi = "10.1103/PhysRevD.110.056040",
    journal = "Phys. Rev. D",
    volume = "110",
    number = "5",
    pages = "056040",
    year = "2024"
}

@article{ATLAS:2023fsd,
    author = "Aad, Georges and others",
    collaboration = "ATLAS",
    title = "{Observation of quantum entanglement with top quarks at the ATLAS detector}",
    eprint = "2311.07288",
    archivePrefix = "arXiv",
    primaryClass = "hep-ex",
    reportNumber = "CERN-EP-2023-230",
    doi = "10.1038/s41586-024-07824-z",
    journal = "Nature",
    volume = "633",
    number = "8030",
    pages = "542--547",
    year = "2024"
}

@article{CMS:2024pts,
    author = "Hayrapetyan, Aram and others",
    collaboration = "CMS",
    title = "{Observation of quantum entanglement in top quark pair production in proton{\textendash}proton collisions at $\sqrt{s} = 13$ TeV}",
    eprint = "2406.03976",
    archivePrefix = "arXiv",
    primaryClass = "hep-ex",
    reportNumber = "CMS-TOP-23-001, CERN-EP-2024-137",
    doi = "10.1088/1361-6633/ad7e4d",
    journal = "Rept. Prog. Phys.",
    volume = "87",
    number = "11",
    pages = "117801",
    year = "2024"
}

@article{Kats:2023zxb,
    author = "Kats, Yevgeny and Uzan, David",
    title = "{Prospects for measuring quark polarization and spin correlations in $b\overline{b }$ and $c\overline{c }$ samples at the LHC}",
    eprint = "2311.08226",
    archivePrefix = "arXiv",
    primaryClass = "hep-ph",
    doi = "10.1007/JHEP03(2024)063",
    journal = "JHEP",
    volume = {2024},
    pages = "063",
    year = "2024",
    number = {03}
}

@article{Galanti:2015pqa,
    author = "Galanti, Mario and Giammanco, Andrea and Grossman, Yuval and Kats, Yevgeny and Stamou, Emmanuel and Zupan, Jure",
    title = "{Heavy baryons as polarimeters at colliders}",
    eprint = "1505.02771",
    archivePrefix = "arXiv",
    primaryClass = "hep-ph",
    reportNumber = "CP3-15-12",
    doi = "10.1007/JHEP11(2015)067",
    journal = "JHEP",
    volume = {2015},
    pages = "067",
    year = "2015",
    number = {11}
}

@article{Falk:1993rf,
    author = "Falk, Adam F. and Peskin, Michael E.",
    title = "{Production, decay, and polarization of excited heavy hadrons}",
    eprint = "hep-ph/9308241",
    archivePrefix = "arXiv",
    reportNumber = "SLAC-PUB-6311, JHU-TIPAC-930019",
    doi = "10.1103/PhysRevD.49.3320",
    journal = "Phys. Rev. D",
    volume = "49",
    pages = "3320--3332",
    year = "1994"
}

@article{vonKuk:2023jfd,
    author = "von Kuk, Rebecca and Michel, Johannes K. L. and Sun, Zhiquan",
    title = "{Transverse momentum distributions of heavy hadrons and polarized heavy quarks}",
    eprint = "2305.15461",
    archivePrefix = "arXiv",
    primaryClass = "hep-ph",
    reportNumber = "MIT-CTP 5537, DESY 23-032, MIT--CTP 5537, DESY 23--032",
    doi = "10.1007/JHEP09(2023)205",
    journal = "JHEP",
    volume = {2023},
    pages = "205",
    year = "2023",
    number = {09}
}

@article{Collins:1992kk,
    author = "Collins, John C.",
    title = "{Fragmentation of transversely polarized quarks probed in transverse momentum distributions}",
    eprint = "hep-ph/9208213",
    archivePrefix = "arXiv",
    reportNumber = "PSU-TH-102",
    doi = "10.1016/0550-3213(93)90262-N",
    journal = "Nucl. Phys. B",
    volume = "396",
    pages = "161--182",
    year = "1993"
}

@article{Boer:1997mf,
    author = "Boer, Daniel and Jakob, R. and Mulders, P. J.",
    title = "{Asymmetries in polarized hadron production in e+ e- annihilation up to order 1/Q}",
    eprint = "hep-ph/9702281",
    archivePrefix = "arXiv",
    reportNumber = "NIKHEF-97-008, VUTH-97-3",
    doi = "10.1016/S0550-3213(97)00456-2",
    journal = "Nucl. Phys. B",
    volume = "504",
    pages = "345--380",
    year = "1997"
}

@article{Boer:2003ya,
    author = "Boer, Dani{\"e}l and Jakob, Rainer and Radici, Marco",
    title = "{Interference fragmentation functions in electron positron annihilation}",
    eprint = "hep-ph/0302232",
    archivePrefix = "arXiv",
    doi = "10.1103/PhysRevD.67.094003",
    journal = "Phys. Rev. D",
    volume = "67",
    pages = "094003",
    year = "2003",
    note = "Erratum: Phys. Rev. D 98, 039902 (2018), doi:10.1103/PhysRevD.98.039902"
}

@article{Metz:2016swz,
    author = "Metz, Andreas and Vossen, Anselm",
    title = "{Parton Fragmentation Functions}",
    eprint = "1607.02521",
    archivePrefix = "arXiv",
    primaryClass = "hep-ex",
    doi = "10.1016/j.ppnp.2016.08.003",
    journal = "Prog. Part. Nucl. Phys.",
    volume = "91",
    pages = "136--202",
    year = "2016"
}

@article{Hatta:2025obw,
    author = "Hatta, Yoshitaka and Schoenleber, Jakob",
    title = "{Probing quantum entanglement with generalized parton distributions at the Electron-Ion Collider}",
    eprint = "2511.04537",
    archivePrefix = "arXiv",
    primaryClass = "hep-ph",
    doi = "10.1103/qdb2-k2nh",
    journal = "Phys. Rev. D",
    volume = "113",
    number = "9",
    pages = "094016",
    year = "2026"
}

@article{Cheng:2025zaw,
    author = "Cheng, Kun and Han, Tao and Trifinopoulos, Sokratis",
    title = "{Quantum information at the electron-ion collider}",
    eprint = "2510.23773",
    archivePrefix = "arXiv",
    primaryClass = "hep-ph",
    reportNumber = "PITT-PACC-2509, CERN-TH-2025-205, MIT-CTP/5943",
    doi = "10.1103/xytf-nhkh",
    journal = "Phys. Rev. D",
    volume = "114",
    number = "5",
    pages = "L051301",
    year = "2026"
}

@article{Cheng:2025cuv,
    author = "Cheng, Kun and Yan, Bin",
    title = "{Bell Inequality Violation of Light Quarks in Dihadron Pair Production at Lepton Colliders}",
    eprint = "2501.03321",
    archivePrefix = "arXiv",
    primaryClass = "hep-ph",
    reportNumber = "PITT-PACC-2414",
    doi = "10.1103/gmqz-v4cl",
    journal = "Phys. Rev. Lett.",
    volume = "135",
    number = "1",
    pages = "011902",
    year = "2025"
}

@article{Bernal:2023jba,
    author = "Bernal, Alexander",
    title = "{Quantum tomography of helicity states for general scattering processes}",
    eprint = "2310.10838",
    archivePrefix = "arXiv",
    primaryClass = "hep-ph",
    doi = "10.1103/PhysRevD.109.116007",
    journal = "Phys. Rev. D",
    volume = "109",
    number = "11",
    pages = "116007",
    year = "2024"
}

@article{Fabbrichesi:2025ywl,
    author = "Fabbrichesi, Marco and Low, Matthew and Marzola, Luca",
    title = "{Trace distance between density matrices: A nifty tool in new-physics searches}",
    eprint = "2501.03311",
    archivePrefix = "arXiv",
    primaryClass = "hep-ph",
    doi = "10.1103/kdmh-3yb4",
    journal = "Phys. Rev. D",
    volume = "112",
    number = "1",
    pages = "013003",
    year = "2025"
}

@misc{Liu:2026dzv,
    author = "Liu, Yu-Xuan and Qi, Wei and He, Luo-Ting and Xiao, Bo-Wen",
    title = "{Polarization, Maximal Concurrence, and Pure States in High-Energy Collisions}",
    eprint = "2604.17756",
    archivePrefix = "arXiv",
    primaryClass = "hep-ph",
    month = "4",
    year = "2026"
}

@misc{Bloss:2026yrf,
    author = "Bloss, Henry and Hobbs, T. J. and McGinnis, Navin",
    title = "{Hadron Structure from the Hierarchy of Quantum Correlations in Deep-Inelastic Scattering}",
    eprint = "2607.28724",
    archivePrefix = "arXiv",
    primaryClass = "hep-ph",
    reportNumber = "ANL-205269",
    month = "7",
    year = "2026"
}

@article{Kullback1951,
  author = {Kullback, S. and Leibler, R. A.},
  title = {On Information and Sufficiency},
  journal = {Ann. Math. Stat.},
  volume = {22},
  number = {1},
  pages = {79--86},
  year = {1951},
  doi = {10.1214/aoms/1177729694}
}

@article{Cowan2011,
  author = {Cowan, Glen and Cranmer, Kyle and Gross, Eilam and Vitells, Ofer},
  title = {Asymptotic formulae for likelihood-based tests of new physics},
  journal = {Eur. Phys. J. C},
  volume = {71},
  pages = {1554},
  year = {2011},
  doi = {10.1140/epjc/s10052-011-1554-0},
  eprint = {1007.1727},
  archivePrefix = {arXiv},
  primaryClass = {physics.data-an},
  note = {Erratum: \href{https://doi.org/10.1140/epjc/s10052-013-2501-z}{Eur. Phys. J. C 73, 2501 (2013)}}
}

@article{Anderle:2021wcy,
    author = "Anderle, Daniele P. and others",
    title = "{Electron-ion collider in China}",
    eprint = "2102.09222",
    archivePrefix = "arXiv",
    primaryClass = "nucl-ex",
    reportNumber = "Frontiers of Physics, Volume 16 Issue (6):64701, 2021",
    doi = "10.1007/s11467-021-1062-0",
    journal = "Front. Phys. (Beijing)",
    volume = "16",
    number = "6",
    pages = "64701",
    year = "2021"
}

@article{Xiao:2026tbs,
    author = "Xiao, Bo-Wen and Zhao, Yuxiang and Zhou, Jian",
    title = "{Physics of the Electron{\textendash}Ion Collider in China}",
    eprint = "2608.11712",
    archivePrefix = "arXiv",
    primaryClass = "hep-ph",
    doi = "10.1016/j.ppnp.2026.104264",
    journal = "Prog. Part. Nucl. Phys.",
    volume = "151",
    pages = "104264",
    year = "2026"
}

@article{Balitsky:2006wa,
    author = "Balitsky, Ian",
    title = "{Quark contribution to the small-x evolution of color dipole}",
    eprint = "hep-ph/0609105",
    archivePrefix = "arXiv",
    reportNumber = "JLAB-THY-06-541",
    doi = "10.1103/PhysRevD.75.014001",
    journal = "Phys. Rev. D",
    volume = "75",
    pages = "014001",
    year = "2007"
}

@article{Kovchegov:2006vj,
    author = "Kovchegov, Yuri V. and Weigert, Heribert",
    title = "{Triumvirate of Running Couplings in Small-x Evolution}",
    eprint = "hep-ph/0609090",
    archivePrefix = "arXiv",
    doi = "10.1016/j.nuclphysa.2006.10.075",
    journal = "Nucl. Phys. A",
    volume = "784",
    pages = "188--226",
    year = "2007"
}

@article{Balitsky:2007feb,
    author = "Balitsky, Ian and Chirilli, Giovanni A.",
    title = "{Next-to-leading order evolution of color dipoles}",
    eprint = "0710.4330",
    archivePrefix = "arXiv",
    primaryClass = "hep-ph",
    reportNumber = "JLAB-THY-07-741",
    doi = "10.1103/PhysRevD.77.014019",
    journal = "Phys. Rev. D",
    volume = "77",
    pages = "014019",
    year = "2008"
}

@article{Iancu:2015vea,
    author = "Iancu, E. and Madrigal, J. D. and Mueller, A. H. and Soyez, G. and Triantafyllopoulos, D. N.",
    title = "{Resumming double logarithms in the QCD evolution of color dipoles}",
    eprint = "1502.05642",
    archivePrefix = "arXiv",
    primaryClass = "hep-ph",
    doi = "10.1016/j.physletb.2015.03.068",
    journal = "Phys. Lett. B",
    volume = "744",
    pages = "293--302",
    year = "2015"
}

@article{Ducloue:2019ezk,
    author = "Duclou{\'e}, B. and Iancu, E. and Mueller, A. H. and Soyez, G. and Triantafyllopoulos, D. N.",
    title = "{Non-linear evolution in QCD at high-energy beyond leading order}",
    eprint = "1902.06637",
    archivePrefix = "arXiv",
    primaryClass = "hep-ph",
    reportNumber = "INT-PUB-19-006",
    doi = "10.1007/JHEP04(2019)081",
    journal = "JHEP",
    volume = "2019",
    number = "04",
    pages = "081",
    year = "2019"
}

@article{Xiang:2022iuh,
    author = "Xiang, Wenchang and Wang, Mengliang and Cai, Yanbing and Zhou, Daicui",
    title = "{Running coupling effect in next-to-leading order Balitsky-Kovchegov evolution equations}",
    doi = "10.1088/1674-1137/ac4ee9",
    journal = "Chin. Phys. C",
    volume = "46",
    number = "5",
    pages = "054104",
    year = "2022"
}

@misc{Dai:2026rgl,
    author = "Dai, Si-Wei and Duan, Haowu and Pang, Long-Gang and Qin, Guang-You and Wei, Shu-Yi and Zhang, Han-Zhong and Zhao, Wenbin",
    title = "{Unbiased Data-Driven Determination of the Nuclear Dipole Amplitude in the Color Glass Condensate}",
    eprint = "2607.27603",
    archivePrefix = "arXiv",
    primaryClass = "hep-ph",
    year = "2026"
}

@article{Mueller:2013wwa,
    author = "Mueller, A. H. and Xiao, Bo-Wen and Yuan, Feng",
    title = "{Sudakov double logarithms resummation in hard processes in the small-x saturation formalism}",
    eprint = "1308.2993",
    archivePrefix = "arXiv",
    primaryClass = "hep-ph",
    doi = "10.1103/PhysRevD.88.114010",
    journal = "Phys. Rev. D",
    volume = "88",
    number = "11",
    pages = "114010",
    year = "2013"
}

@article{Zhou:2016tfe,
    author = "Zhou, Jian",
    title = "{The evolution of the small x gluon TMD}",
    eprint = "1603.07426",
    archivePrefix = "arXiv",
    primaryClass = "hep-ph",
    doi = "10.1007/JHEP06(2016)151",
    journal = "JHEP",
    volume = "2016",
    number = "06",
    pages = "151",
    year = "2016"
}

@article{Zheng:2018ssm,
    author = "Zheng, Liang and Aschenauer, E. C. and Lee, J. H. and Xiao, Bo-Wen and Yin, Zhong-Bao",
    title = "{Accessing the gluon Sivers function at a future electron-ion collider}",
    eprint = "1805.05290",
    archivePrefix = "arXiv",
    primaryClass = "hep-ph",
    doi = "10.1103/PhysRevD.98.034011",
    journal = "Phys. Rev. D",
    volume = "98",
    number = "3",
    pages = "034011",
    year = "2018"
}

@article{Hatta:2020bgy,
    author = "Hatta, Yoshitaka and Xiao, Bo-Wen and Yuan, Feng and Zhou, Jian",
    title = "{Anisotropy in Dijet Production in Exclusive and Inclusive Processes}",
    eprint = "2010.10774",
    archivePrefix = "arXiv",
    primaryClass = "hep-ph",
    doi = "10.1103/PhysRevLett.126.142001",
    journal = "Phys. Rev. Lett.",
    volume = "126",
    number = "14",
    pages = "142001",
    year = "2021"
}

@article{Kang:2020xgk,
    author = "Kang, Zhong-Bo and Reiten, Jared and Shao, Ding Yu and Terry, John",
    title = "{QCD evolution of the gluon Sivers function in heavy flavor dijet production at the Electron-Ion Collider}",
    eprint = "2012.01756",
    archivePrefix = "arXiv",
    primaryClass = "hep-ph",
    doi = "10.1007/JHEP05(2021)286",
    journal = "JHEP",
    volume = "2021",
    number = "05",
    pages = "286",
    year = "2021"
}

@article{Zhao:2021kae,
    author = "Zhao, Ye-Yin and Xu, Ming-Mei and Chen, Li-Zhu and Zhang, Dong-Hai and Wu, Yuan-Fang",
    title = "{Suppressions of dijet azimuthal correlations in the future EIC}",
    eprint = "2105.08818",
    archivePrefix = "arXiv",
    primaryClass = "hep-ph",
    doi = "10.1103/PhysRevD.104.114032",
    journal = "Phys. Rev. D",
    volume = "104",
    number = "11",
    pages = "114032",
    year = "2021"
}

@article{Hatta:2021jcd,
    author = "Hatta, Yoshitaka and Xiao, Bo-Wen and Yuan, Feng and Zhou, Jian",
    title = "{Azimuthal angular asymmetry of soft gluon radiation in jet production}",
    eprint = "2106.05307",
    archivePrefix = "arXiv",
    primaryClass = "hep-ph",
    doi = "10.1103/PhysRevD.104.054037",
    journal = "Phys. Rev. D",
    volume = "104",
    number = "5",
    pages = "054037",
    year = "2021"
}

@article{Dong:2022xbd,
    author = "Dong, Xin and Ji, Yuanjing and Kelsey, Matthew and Radhakrishnan, Sooraj and Sichtermann, Ernst and Zhao, Yuxiang",
    title = "{Probing gluon TMDs with reconstructed and tagged heavy flavor hadron pairs at the EIC}",
    eprint = "2210.08609",
    archivePrefix = "arXiv",
    primaryClass = "hep-ph",
    doi = "10.1103/PhysRevD.107.074022",
    journal = "Phys. Rev. D",
    volume = "107",
    number = "7",
    pages = "074022",
    year = "2023"
}

@misc{Lin:2026typ,
    author = "Lin, Shuo and Zhou, Jian",
    title = "{Evading Sudakov Dilution in Gluon Tomography}",
    eprint = "2609.02280",
    archivePrefix = "arXiv",
    primaryClass = "hep-ph",
    month = "9",
    year = "2026"
}

@misc{Marquet:2025jdr,
    author = "Marquet, Cyrille and Shi, Yu and Xiao, Bo-Wen",
    title = "{Unified Resummation of Soft Gluon Radiation in Heavy Meson Pair Photoproduction}",
    eprint = "2510.18949",
    archivePrefix = "arXiv",
    primaryClass = "hep-ph",
    month = "10",
    year = "2025"
}

@article{Chen:1994ar,
    author = "Chen, Kun and Goldstein, Gary R. and Jaffe, R. L. and Ji, Xiangdong",
    title = "{Probing quark fragmentation functions for spin 1/2 baryon production in unpolarized e+ e- annihilation}",
    eprint = "hep-ph/9410337",
    archivePrefix = "arXiv",
    reportNumber = "MIT-CTP-2365",
    doi = "10.1016/0550-3213(95)00193-V",
    journal = "Nucl. Phys. B",
    volume = "445",
    pages = "380--396",
    year = "1995"
}

@article{Zhang:2023ugf,
    author = "Zhang, Hao-Cheng and Wei, Shu-Yi",
    title = "{Probing the longitudinal spin transfer via dihadron polarization correlations in unpolarized $e^+e^-$ and pp collisions}",
    eprint = "2301.04096",
    archivePrefix = "arXiv",
    primaryClass = "hep-ph",
    doi = "10.1016/j.physletb.2023.137821",
    journal = "Phys. Lett. B",
    volume = "839",
    pages = "137821",
    year = "2023"
}

@article{Yang:2024kjn,
    author = "Yang, Lei and Song, Yu-Kun and Wei, Shu-Yi",
    title = "{Transverse spin correlation of back-to-back dihadrons in unpolarized collisions}",
    eprint = "2410.20917",
    archivePrefix = "arXiv",
    primaryClass = "hep-ph",
    doi = "10.1103/PhysRevD.111.054035",
    journal = "Phys. Rev. D",
    volume = "111",
    number = "5",
    pages = "054035",
    year = "2025"
}

@article{Xi:2025feb,
    author = "Xi, Xue-Qi and Chen, Kai-Bao and Tong, Xuan-Bo and Wei, Shu-Yi and Wu, Jing",
    title = "{Helicity correlation of dihadron in current and target fragmentation regions of unpolarized SIDIS}",
    eprint = "2510.06103",
    archivePrefix = "arXiv",
    primaryClass = "hep-ph",
    doi = "10.1140/epjc/s10052-025-15193-8",
    journal = "Eur. Phys. J. C",
    volume = "85",
    number = "12",
    pages = "1458",
    year = "2025"
}

@article{Zhao:2024usu,
    author = "Zhao, Xiaoyan and Liang, Zuo-tang and Liu, Tianbo and Zhou, Ya-jin",
    title = "{Suppression of Spin Transfer to {\ensuremath{\Lambda}} Hyperon in Deep-Inelastic Scattering}",
    eprint = "2411.06205",
    archivePrefix = "arXiv",
    primaryClass = "hep-ph",
    doi = "10.1103/1nd6-9nw9",
    journal = "Phys. Rev. Lett.",
    volume = "134",
    number = "23",
    pages = "231901",
    year = "2025"
}

@article{Wu:2024mtj,
    author = "Wu, Sihao and Qian, Chen and Yang, Yang-Guang and Wang, Qun",
    title = "{Generalized Quantum Measurement in Spin-Correlated Hyperon-Antihyperon Decays}",
    eprint = "2402.16574",
    archivePrefix = "arXiv",
    primaryClass = "hep-ph",
    doi = "10.1088/0256-307X/41/11/110301",
    journal = "Chin. Phys. Lett.",
    volume = "41",
    number = "11",
    pages = "110301",
    year = "2024"
}

@article{Schonning:2023mge,
    author = {Sch{\"o}nning, Karin and Batozskaya, Varvara and Adlarson, Patrik and Zhou, Xiaorong},
    title = "{Production and decay of polarized hyperon-antihyperon pairs}",
    eprint = "2302.13071",
    archivePrefix = "arXiv",
    primaryClass = "hep-ph",
    doi = "10.1088/1674-1137/acc790",
    journal = "Chin. Phys. C",
    volume = "47",
    number = "5",
    pages = "052002",
    year = "2023"
}

@article{Bernreuther:2015yna,
    author = "Bernreuther, Werner and Heisler, Dennis and Si, Zong-Guo",
    title = "{A set of top quark spin correlation and polarization observables for the LHC: Standard Model predictions and new physics contributions}",
    eprint = "1508.05271",
    archivePrefix = "arXiv",
    primaryClass = "hep-ph",
    reportNumber = "TTK-15-16",
    doi = "10.1007/JHEP12(2015)026",
    journal = "JHEP",
    volume = "2015",
    number = "12",
    pages = "026",
    year = "2015"
}

@article{Bernreuther:2024ltu,
    author = "Bernreuther, Werner and Chen, Long and Si, Zong-Guo",
    title = "{Binned top quark spin correlation and polarization observables for the LHC at 13.6~TeV}",
    eprint = "2403.04371",
    archivePrefix = "arXiv",
    primaryClass = "hep-ph",
    reportNumber = "TTK-24-11",
    doi = "10.1103/PhysRevD.109.116016",
    journal = "Phys. Rev. D",
    volume = "109",
    number = "11",
    pages = "116016",
    year = "2024"
}

@article{Lin:2025eci,
    author = "Lin, Shi-Jia and Liu, Ming-Jun and Shao, Ding Yu and Wei, Shu-Yi",
    title = "{Spin correlations and Bell nonlocality in $ \Lambda \overline{\Lambda} $ pair production from e$^{+}$e$^{-}$ collisions with a thrust cut}",
    eprint = "2507.15387",
    archivePrefix = "arXiv",
    primaryClass = "hep-ph",
    doi = "10.1007/JHEP11(2025)082",
    journal = "JHEP",
    volume = "2025",
    number = "11",
    pages = "082",
    year = "2025"
}

@misc{vonKuk:2025kbv,
    author = "von Kuk, Rebecca and Lee, Kyle and Michel, Johannes K. L. and Sun, Zhiquan",
    title = "{Towards a Quantum Information Theory of Hadronization: Dihadron Fragmentation and Neutral Polarization in Heavy Baryons}",
    eprint = "2503.22607",
    archivePrefix = "arXiv",
    primaryClass = "hep-ph",
    reportNumber = "MIT-CTP 5818, Nikhef 2024-020, DESY-25-034",
    month = "3",
    year = "2025"
}

@article{Cheung:2024qvw,
  author = {V. Cheung and Z.-B. Kang and F. Salazar and R. Vogt},
  title = {{Direct quarkonium production in DIS from a joint CGC and NRQCD framework}},
  journal = {Phys. Rev. D},
  volume = {110},
  number = {9},
  pages = {094039},
  year = {2024},
  doi = {10.1103/PhysRevD.110.094039},
  eprint = {2409.04080},
  archivePrefix = {arXiv}
}

\end{document}